\documentclass[preprint,12pt]{elsarticle}

\usepackage{amssymb}

\usepackage{lineno}

\biboptions{sort&compress}

\usepackage[figuresright]{rotating}

\usepackage{todonotes}
\usepackage{amsmath}
\usepackage{graphicx}
\usepackage{subcaption}
\usepackage{changepage}
\usepackage{hyperref}
\usepackage[capitalise]{cleveref}
\usepackage{float}
\usepackage{booktabs} 

\newboolean{SHOWCOMMENTS}
\setboolean{SHOWCOMMENTS}{true}

\ifthenelse{\boolean{SHOWCOMMENTS}}{
\newcommand{\tdHZ}[1]
{\todo[color=green!25,inline]{\footnotesize{\bf Henrik:} #1}}
\newcommand{\scc}[1]
{\todo[color=red!25,inline]{\footnotesize{\bf Seba:} #1}}
}{
\newcommand{\tdHZ}[1]{}
\newcommand{\scc}[1]{}
}

\renewcommand{\d}[1]{\textnormal{d}#1}

\begin{document}

\begin{frontmatter}





\title{Risk-mediated dynamic regulation of effective contacts de-synchronizes outbreaks in metapopulation epidemic models}


\author[a]{Henrik Zunker\fnref{ca}}
\ead{henrik.zunker@dlr.de}
\author[b,c]{Philipp Dönges}
\author[a]{Patrick Lenz}
\author[b,c]{Seba Contreras\fnref{la}}
\ead{seba.contreras@ds.mpg.de}
\author[a,d]{Martin J. Kühn\fnref{la}}
\ead{martin.kuehn@dlr.de}

\fntext[ca]{Corresponding author}
\fntext[la]{Shared last author in alphabetic order}

\address[a]{ Institute of Software Technology, Department of High-Performance Computing,
German Aerospace Center, Cologne, Germany}
\address[b]{ Max Planck Institute for Dynamics and Self-Organization, Göttingen, Germany,}
\address[c]{ Institute for the Dynamics of Complex Systems, University of Göttingen, Göttingen, Germany}
\address[d]{ Life and Medical Sciences Institute and Bonn Center for Mathematical Life Sciences, University of Bonn, Bonn, Germany}

\begin{abstract}
Metapopulation epidemic models help capture the spatial dimension of infectious disease spread by dividing heterogeneous populations into separate but interconnected communities, represented by nodes in a network.
In the event of an epidemic, an important research question is, to what degree spatial information (i.e., regional or national) is relevant for mitigation and (local) policymakers.
This study investigates the impact of different levels of information on nationwide epidemic outcomes, modeling the reaction to the measured hazard 
as a feedback loop reducing contact rates in a metapopulation model based on ordinary differential equations (ODEs). 
Using COVID-19 and high-resolution mobility data for Germany of 2020 as a case study, we found two markedly different regimes depending on the maximum contact reduction $\psi_{\rm max}$: mitigation and suppression. In the regime of (modest) mitigation, gradually increasing $\psi_{\rm max}$ from zero to moderate levels delayed and spread out the onset of infection waves while gradually reducing the peak values. This effect was more pronounced when the contribution of regional information was low relative to national data.
In the suppression regime, the feedback-induced contact reduction is strong enough to extinguish local outbreaks and decrease the mean and variance of the peak day distribution, thus regional information was more important. 
When suppression or elimination is impossible, ensuring local epidemics are desynchronized helps to avoid hospitalization or intensive care bottlenecks by reallocating resources from less-affected areas. 
\end{abstract}


\end{frontmatter}


\section{Introduction}
\label{sec::introduction}

Infectious diseases like COVID-19 may spread from person to person in contact networks, should one of the interacting individuals be infectious, the other susceptible to the pathogen, and the contact close enough to be contagious. Classic ordinary- or integro-differential equation-based compartmental models (ODE or IDE) offer reasonable results and qualitative agreement in mean with the overall dynamics of the disease in homogeneous populations~\cite{10.3389/fphy.2022.842180,Vasilis2024,ploetzke2024,Wendler2024IDE}. However, one restrictive hypothesis in these models is that of homogeneous mixing in populations; thus, using them typically overlooks the spatial dimension of an outbreak. In large territories with diverse population distributions, more complex approaches are required.

A straightforward extension of compartmental models of disease spread to incorporate the spatial dimension of an outbreak is the use of metapopulation models~\cite{pei_differential_2020,kuhn_assessment_2021,kuhn_regional_2022,chen_compliance_2021,levin_effects_2021,liu_modelling_2022,zunker_novel_2024,contreras2020multi}. While there are different approaches among the prior, in~\cite{kuhn_assessment_2021,kuhn_regional_2022,zunker_novel_2024,contreras2020multi}, spatially separated communities (where homogeneous mixing is assumed) are represented by nodes, and the exchange between them (e.g., commuters or travelers) as edges connecting them. Nodes in these networks have dynamics, which are represented by decoupled ODEs. This is, in contrast to~\cite{pei_differential_2020,chen_compliance_2021,levin_effects_2021,liu_modelling_2022} where the local dynamics are coupled through the force of infection and fluxes to the ODEs representing other nodes that are connected to them in a formalism equivalent to that of including age-stratification (see, e.g., \cite{bauer_relaxing_2021}). Both approaches are equally valid metapopulation models that can represent an epidemic's spatio-temporal dynamics. 

As the disease spreads within and between nodes, information about the outbreak reaches the general population and local governments. Based on the local progression of the disease and the quantified and perceived risk, individuals and policymakers might attempt to reduce the effective contacts (i.e., those that bear the potential to infect others), thereby affecting the spreading dynamics of the disease itself in a complex feedback loop (see, e.g., the foundational work of d'Onofrio and coauthors~\cite{donofrio2009information,banerjee2023spatio,banerjee2024behavior,della2021volatile} and references therein) that can also be observed in longitudinal surveys, e.g., the German COSMO study~\cite{Betsch_Wieler_Bosnjak_Ramharter_Stollorz_Omer_Korn_Sprengholz_Felgendreff_Eitze_Schmid_2020,BETSCH20201255}. The incorporation of a feedback mechanism representing human behavior not only induces complex and chaotic dynamics in epidemic models~\cite{donofrio2009information,wagner2023societal, stollenwerk2022seasonally, Zozmann2024}, but also fixes unrealistically high surges in SIR-like dynamics in fixed-rate epidemic models and stabilizes the system at a lower endemic equilibrium or steady state dynamics~\cite{10.3389/fphy.2022.842180,contreras2023emergency}. In a metapopulation framework, where information (and thus hazard) can be measured at different spatial scales (local, regional, and national), how should local authorities weigh these risks when designing interventions?

In this work, we answer this question using a graph-based metapopulation compartmental model, including high-resolution travel mechanisms between nodes and a feedback loop to dynamically adapt the effective contacts based on different levels of spatial information. Using parameters representing the COVID-19 pandemic in Germany in autumn 2020 as a case study, we investigate the effect of dynamic interventions and different degrees of spatial information.

\section{Methods}
\subsection{Disease spread model}
In this section, we present a comprehensive mathematical model that can be used to simulate the spread of infectious diseases, taking into account human behavior and non-pharmaceutical interventions (NPIs) in spatial resolution from local through regional to nation- or system-wide. The fundamental element of our approach is an ODE-based SECIR-type model that categorizes individuals into different infection states and age groups. ODE models, in general, are characterized by their computational efficiency at the population level, especially compared to agent-based models such as~\cite{kerr_covasim_2021,muller_predicting_2021,KKN24}. However, they are often unable to capture the spatial heterogeneity of disease spread. To overcome this limitation, we use a hybrid graph-ODE metapopulation approach ~\cite{kuhn_assessment_2021,contreras2020multi}. In this approach, different geographic regions are represented as nodes in a graph, where edges represent the (regular) mobility and interaction between these nodes.

In addition to spatial resolution, our model incorporates a feedback mechanism that represents the (potentially delayed) mitigation response that society as a whole has when facing an epidemic. This mechanism is based on the formalism proposed by d'Onofrio and Manfredi~\cite{donofrio2009information, wagner2023societal}, adapted for complex models with age structure \cite{10.3389/fphy.2022.842180}. In particular, the effective spreading rate of the disease is reduced when the overall infection risk perception among the population increases. This perception of risk is based on (or is directly proportional to) the trends of ICU occupancy---which is the information individuals receive. Combining this formalism with the spatial nature of our hybrid graph-ODE model allows us to combine different sources of information that are generated in different spatial units (e.g., local, regional, and national). 

The following subsections explain each of the components in more detail.

\subsubsection{Hybrid graph-ODE approach}
\label{sec::graph}

To incorporate spatial resolution into our model, we utilize the graph approach as proposed in~\cite{kuhn_assessment_2021}. There, each geographic entity is represented as a node within a multigraph (cf.~\cref{fig:ode_model_overview}a for a symbolic scheme using the example of Germany). Given that our model is age-stratified and comprises different disease states, the number of edges connecting a pair of nodes is given by the product of the number of age groups and number of compartments in the model.

We implement two daily exchanges of commuters that occur instantaneously, at the beginning of the day and after half a day. With this approach, commuters can change their infection state during the half-day stay. 

The number of individuals commuting between nodes is derived from the data provided in~\cite{bmas_pendlerverflechtungen_2020, kuhn_vorlaufige_2022}. Since the data contains numerous values close to or equal to zero, we introduce a threshold to exclude weakly connected node pairs to improve the computational efficiency. After filtering, approximately 40,000 edges remain for the 400 counties in Germany.

\subsubsection{ODE-SECIR-type model}\label{sec::ode_model}

\begin{figure}[ht!]%
\hspace*{-2.5 cm}
        \centering
       	\includegraphics[width = 180mm]{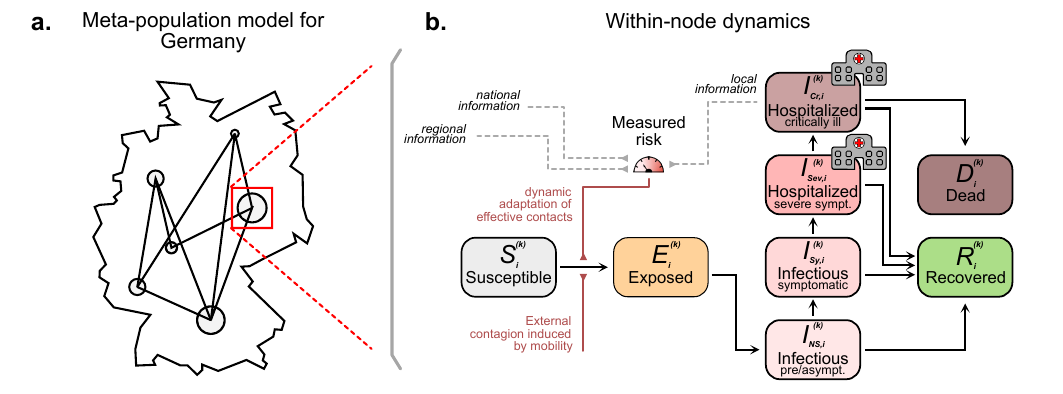}
       	\caption{\textbf{Overview of the SECIR-type model integrated in a graph-based approach~\cite{kuhn_assessment_2021}, with additional modeling of behavior.} Behavioral adaptations can be resulting from the interaction between risk perception and voluntary health-protective actions~\cite{10.3389/fphy.2022.842180} or from ICU-based dynamic NPIs~\cite{kuhn_assessment_2021}. The arrow that is impacted by the integration of the behavior is marked in red. The ICU infection state is represented by the critically ill, hospitalized infection state, which plays a  crucial role in the calculation of the integrated risk for the modeling of risk-mediated regulation.}
       	\label{fig:ode_model_overview}%
\end{figure}

Each of the nodes in the graph (corresponding to different regions in the metapopulation) has its own disease-spread dynamics, which we represent using an ODE-based SECIR-type model~\cite{kuhn_assessment_2021} with particular consideration of hospitalization and intensive care treatment. 
This model structures the course of the disease into the infection states:
\textit{Susceptible} ($S$), \textit{Exposed} ($E$), \textit{Nonsymptomatic Infectious} ($I_{NS}$ or termed as Carrier, C), and potentially progresses through \textit{Symptomatic Infectious} ($I_{Sy}$), \textit{Infected Severe} ($I_{Sev}$), understood as hospitalized, \textit{Infected Critical} ($I_{Cr}$, understood as ICU), to \textit{Dead} ($D$). Starting from the nonsymptomatic state, individuals can always recover ($R$) without ever presenting symptoms. Additionally, our model is stratified into age groups (subindex $i$) that interact with each other according to a contact matrix assumed to be uniform across the populations. The age-resolved equations for an arbitrary node $k$ of the metapopulation are given by:

\begin{align}
\begin{aligned}
    \frac{\d S^{(k)}_{i}}{\d t} &= - S^{(k)}_{i}\rho^{(k)}_{i}\,\sum_{j=1}^n \phi^{(k)}_{i,j} \frac{{\xi^{(k)}_{I_{NS},j}}I^{(k)}_{NS,j} + {\xi^{(k)}_{I_{Sy},j}}I^{(k)}_{Sy,j}}{{N^{(k)}_j- D^{(k)}_j}},\\
    \frac{\d E^{(k)}_{i}}{\d t} &= S^{(k)}_{i}\rho^{(k)}_{i}\,\sum_{j=1}^n \phi^{(k)}_{i,j} \frac{{\xi^{(k)}_{I_{NS},j}}I^{(k)}_{NS,j} + {\xi^{(k)}_{I_{Sy},j}}I^{(k)}_{Sy,j}}{{N^{(k)}_j- D^{(k)}_j}}
    - \frac{E^{(k)}_{i}}{T^{(k)}_{E_{i}}} ,\\
    \frac{\d I^{(k)}_{NS,i}}{\d t} &= \frac{E^{(k)}_{i}}{T^{(k)}_{E_{i}}} - \frac{I^{(k)}_{NS,i}}{T^{(k)}_{I_{NS,i}}} ,\\
    \frac{\d I^{(k)}_{Sy,i}}{\d t} &= \frac{\mu_{I_{NS,i}}^{I_{Sy,i}, (k)}}{T^{(k)}_{I_{NS,i}}} I^{(k)}_{NS,i}-\frac{I^{(k)}_{Sy,i}}{T^{(k)}_{I_{Sy,i}}} ,\\
    \frac{\d I^{(k)}_{Sev,i}}{\d t} &= \frac{\mu_{I_{Sy,i}}^{I_{Sev,i}, (k)}}{T^{(k)}_{I_{Sy,i}}}I^{(k)}_{Sy,i} - \frac{I^{(k)}_{Sev,i}}{T^{(k)}_{I_{Sev,i}}} ,\\ 
    \frac{\d I^{(k)}_{Cr,i}}{\d t} &= \frac{\mu_{I_{Sev,i}}^{I_{Cr,i},(k)}}{T^{(k)}_{I_{Sev,i}}} I^{(k)}_{Sev,i} - \frac{I^{(k)}_{Cr,i}}{T^{(k)}_{I_{Cr,i}}},\\
    \frac{\d D^{(k)}_{i}}{\d t} &= \frac{\mu_{I_{Cr,i}}^{D_{i},(k)}}{T^{(k)}_{I_{Cr,i}}} I^{(k)}_{Cr,i},\\
    \frac{\d R^{(k)}_{i}}{\d t} &=  \frac{1 - \mu_{I_{NS,i}}^{I_{Sy,i},(k)}}{T^{(k)}_{I_{NS,i}}} I^{(k)}_{NS,i} + \frac{1 - \mu_{I_{Sy,i}}^{I_{Sev,i},(k)}}{T^{(k)}_{I_{Sy,i}}} I^{(k)}_{Sy,i}\\
    &\hspace{1.5cm}+ \frac{1 - \mu_{I_{Sev,i}}^{I_{Cr,i},(k)}}{T^{(k)}_{I_{Sev,i}}} I^{(k)}_{Sev,i} + \frac{1 - \mu_{I_{Cr,i}}^{D_{i},(k)}}{T^{(k)}_{I_{Cr,i}}} I^{(k)}_{Cr,i},
    \label{eq:ODESECIR}
\end{aligned}
\end{align}
where $\phi^{(k)}_{i,j}(t)$ represents the effective contact rate between any age groups $i$ and $j$ in region $k$. For the sake of simplicity, we omitted the explicit dependence on $t$ in~\eqref{eq:ODESECIR}.
The contact rate for Germany is obtained from studies presented in~\cite{prem_projecting_2017, fumanelli_inferring_2012}. All parameters are explained in Table~\ref{tab:ODE_params}, and~\cref{fig:ode_model_overview} represents an overview of the dynamics in the metapopulation (which is exemplified using a sketch of Germany).

\begin{table}[ht]
    \begin{adjustwidth}{-0.5in}{0in}
    \centering
    \caption{\textbf{Parameters used in Model~\eqref{eq:ODESECIR}.} For clarity the superindex $k$ (representing the region in the metacommunity) has been excluded.}
    \begin{tabular}{ll}
        \toprule
        \textbf{Parameter} & \textbf{Description} \\ \midrule
        $N_i$ & Population size of age group $i$. \\ 
        $\rho_i(t)$ & Age-dependent transmission rate of the disease. \\ 
        ${\phi}_{i,j}(t)$ & Average number of effective contacts between age groups $i$ and $j$ per day. \\ 
        $\xi_{I_{NS},i}(t)$ & Proportion of nonisolated asymptomatic infectious individuals of age group $i$. \\ 
        $\xi_{I_{Sy},i}(t)$ & Proportion of nonisolated symptomatic infectious individuals of age group $i$. \\ 
        $T_{z}$ & Average duration in days an individual stay in compartment $z$.\\ 
        $\mu_{z_1}^{z_2}$ & Transition probability from compartment $z_1$ to $z_2$. \\ 
        \bottomrule
    \end{tabular}
    \label{tab:ODE_params}
\end{adjustwidth}
\end{table}

\subsubsection{Feedback loop for the risk-dynamic regulation of effective contacts}\label{sec::behavior}

To calculate the dynamic regulation of the effective contacts based on the ICU occupancy, we first calculate the perceived risk $H:\mathbf{R}^+_0\to\mathbf{R}_0^{+}$, which, for simplicity, we assume to be uniform across age groups. Based on~\cite{10.3389/fphy.2022.842180}, the perceived risk is convolved with a Gamma kernel $\mathcal{G}_{a,b}$ representing memory and information processing, including past values of the ICU in the current perception of risk and weighing them according to their recency. 
The Gamma function is defined using shape parameter $a$ and scaling parameter $b$. 
Therefore, the perceived risk for a local model without spatial stratification is defined by
\begin{align}\label{eq:perceived_risk}
    H(t) = \int_{-\infty}^{t} \frac{\sum_{i} I_{Cr,i}(\tau)}{C_{ICU}}   \  \mathcal{G}_{a,b}(t-\tau)\mathrm{d}\tau,
\end{align}

where $C_{ICU}$ represents the nominal ICU capacity, which serves as a critical threshold for triggering governmental intervention measures. Several feasible mitigation policies are implemented to reduce transmission when getting close to this threshold. In our model, risk perception $H$ is updated at the beginning of each day, reflecting realistic possible responses. Note that we here build upon and borrow nomenclature from prior studies on the self-regulation of effective contacts. However, the perceived risk $H$ (see Eq.~\eqref{eq:perceived_risk}) can be understood as a measured hazard that triggers targeted NPIs, which reduce the number of effective contacts.

The functional shape of the contact reductions in region $k$ triggered by risk perception at a given time $\psi^{(k)}_{self}(t)$ is defined as in \cite{10.3389/fphy.2022.842180}, using a soft-plus function, i.e., a function with a linear growth stage following by a saturation level, with a smooth transition:
\begin{align}\label{eq:self_reg_contacts}
    \psi^{(k)}_{self}(t) = (\psi_{max} - \psi_{min}) \cdot \epsilon \cdot \log\left(\exp\left(\frac{H(t)}{\epsilon}\right) + 1\right),
\end{align}
where $\epsilon$ is a curvature parameter, and $\psi_{min},\,\psi_{max}$ represent the bounds on the extent to which (self-)regulation can influence the contact rate. The final expression for the effective contact rate between age groups $i$ and $j$ in region $k$ is:
\begin{align}\label{eq:contacts_update}
     {\phi}_{i,j}^{(k)}(t) = (1 - \psi^{(k)}_{self}(t)) \cdot \phi_{i,j,0}^{(k)},
\end{align} 
where $\phi_{i,j,0}^{(k)}$ denotes a constant initial contact rate.

\subsubsection{Spatial Integration of Behavioral Feedback}

So far, we have calculated a population-weighted average of the local-level risks for each node in the graph (i.e., regions in the metapopulation approach). Let $P^{(k)}=\sum_i N^{(k)}_i$ denote the total population of node $k$. The perceived risk for a higher spatial level, such as a federal state or an entire country, is then computed by
\begin{align}\label{eq:aggregated_risk}
    H_{\textit{agg}}^{(k)}(t) = \frac{\sum_{l \in \mathcal{R}^{(k)}} P^{(l)} \cdot H^{(l)}(t)}{\sum_{l \in \mathcal{R}^{(k)}} P^{(l)}},
\end{align}
where $\mathcal{R}^{(k)}$ describes the set of all local units within the considered aggregation size (e.g., federal state or national) defined by node $k$.

To integrate the aggregated risk into the feedback approach, we introduce weighting factors for each spatial level, allowing us to calculate a composite risk that reflects the influence of risks at the local, regional, and national levels.

Let $w_l,\,w_r,\,w_n \in [0,1]$ represent the weighting factors for the local, regional, and national levels, respectively. These weighting factors are defined such that their sum equals one. Using these weighting factors, we compute the adjusted risk $\widetilde{H}^{(k)}$ for a region  corresponding to the node with index $k$ as a linear combination of the risks at different spatial scales,
\begin{align}\label{eq:total_perceived_risk}
    \widetilde{H}^{(k)}(t) = w_l H^{(k)}(t) + w_r \cdot H_{agg,r}^{(k)}(t) + w_{n} \cdot H_{agg,n}(t).
\end{align}
Here, $H_{agg,r}^{(k)}$ represents the risk aggregated at regional level for the region corresponding to the node with index $k$, while $H_{agg,n}$ denotes the nationally aggregated risk.

This approach allows us to account for the varying impact of risks across different spatial scales, ensuring that the overall risk used in the behavioral feedback mechanism accurately reflects the perceived risk or quantified hazard. By adjusting the weighting factors, we can model different scenarios where, for instance, local dynamics may be more influential than national dynamics.

This adjusted risk $\widetilde{H}^{(k)}$ can then be used in Eq~\eqref{eq:self_reg_contacts} to determine the degree of response implemented at different spatial levels, whether through voluntary behavioral adaptations or through the application of dynamic NPIs.

\subsection{Definition of regional proximity}\label{sec::regional}

As provided in Eq.~\eqref{eq:total_perceived_risk}, we will analyze three different spatial levels of information that contribute to the risk: local (node, county), regional (local proximity measure or federal state), and national (whole graph). While the unit and the whole are well-defined, the definition of the region has some caveats. On the one hand, definitions based on political boundaries (e.g., counties in a particular federal state) may not reflect the heterogeneity in the size of political units or synthetically introduce barriers that are nonexistent to disease spread. On the other hand, distance---or proximity---based measures may not have political meaning and are hard to define in a metapopulation formalization of spatial units, as it is unclear how to include "fractions" of regions, but they are more realistic for disease spread. We define regional proximity, by inclusion of counties whose centroids fall within a certain radius from the county of interest.

\begin{figure}[ht!]
    \centering
    \begin{subfigure}[b]{0.45\textwidth}
        \centering
        \includegraphics[width=0.6\textwidth]{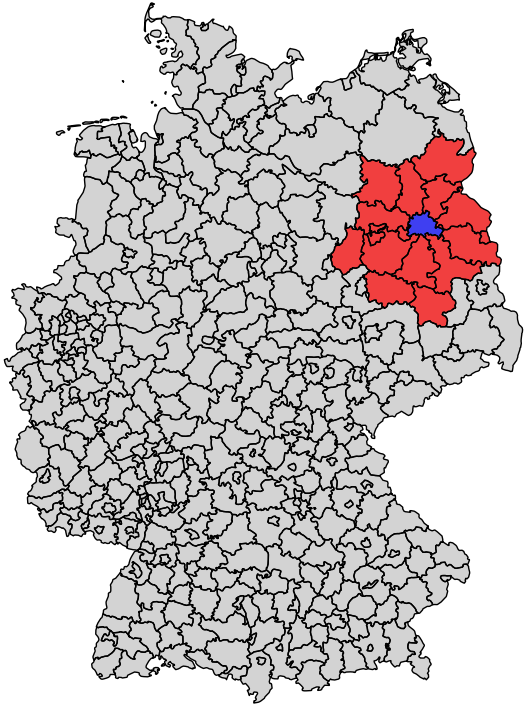}
        \caption{Regional proximity of Cologne.}
    \end{subfigure}
    \hfill
    \begin{subfigure}[b]{0.45\textwidth}
        \centering
        \includegraphics[width=0.6\textwidth]{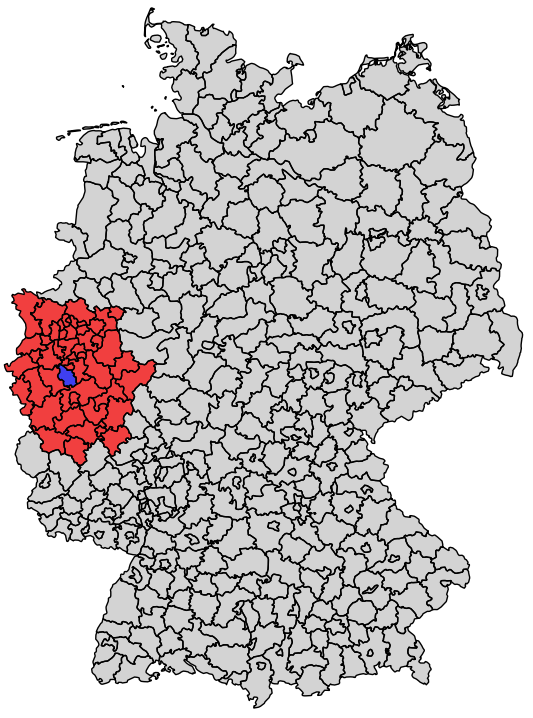}
        \caption{Regional proximity of Berlin.}
    \end{subfigure}
    \caption{{Regional proximity of counties Cologne and Berlin.} (a) Cologne and (b) Berlin; using the national average radius of approximately 103 km in Germany. The selected county is shown in purple. Counties with centroid located within 103 km radius, are considered to be in regional proximity and are shown in red, all other counties are displayed in gray.}
    \label{fig:random_counties_comparison}
\end{figure}

Using Germany as an example, we first calculate the centroid of each county. Then, we determine the distances between each county and the other counties within the same federal state. From these distances, we calculate the weighted mean distance for each of the 16 federal states in Germany, adjusting the measure based on the number of counties in each state. This results in a national average radius of approximately 103 km. Subsequently, we define regionally close counties by integration of all counties within this radius. This is then used for the calculation of the risk metric $H$ in Eq~\eqref{eq:aggregated_risk}. \cref{fig:random_counties_comparison} illustrates the regional proximity of two counties, Cologne and Berlin. 

\subsection{Definition of epidemic waves}\label{sec::wave}

While in common language, rather imprecise time windows and beginnings were used to subdivide the COVID-19 pandemic into different waves, we need a clear definition for an epidemic wave to analyze the outcomes of various scenarios. Here, we define the timing of an epidemic wave as the time point between three consecutive increases and three consecutive declines in the observable of interest (i.e., incidence or ICU occupancy). This allows us to rule out "border effects" that would misclassify the initial or final point of monotonic epidemic curves as waves.

\newpage
\section{Results}\label{sec::Results}

In this section, we numerically examine how risk-mediated dynamic regulation of effective contacts influences the spatial spread of epidemics. Additionally, we consider the impact of varying levels of locality in the information used to shape risk perception or quantification. Our study focuses on a scenario that reflects the dynamics of SARS-CoV-2 in Germany during autumn 2020, with October 1st as the start date, a time before vaccines were introduced. The model is divided into six age groups, with age-resolved parameters (see \cref{tab:covid_age_dep_parameters,tab:covid_parameters})
and initialization based on~\cite{kuhn_assessment_2021} using official reported data~\cite{Robert_Koch-Institut_SARS-CoV-2_Infektionen_in_2025, DIVI22, RKIVACC}. To assess the robustness of our results, we also tested three additional settings with starting dates September 1st, November 1st, and December 1st, 2020 (see supplementary materials).

\begin{table}[htbp]
    \centering
    \caption{\textbf{Age-dependent epidemiological parameters.} Parameters are based on~\cite{kuhn_assessment_2021} for the SARS-CoV-2 dynamics in Germany, autumn 2020.}
    \begin{adjustwidth}{-0.9in}{0in}
    \begin{tabular}{lp{5.3cm}lccccccc}
        \toprule
        Parameter & Description & Units & 0-4 & 5-14 & 15-34 & 35-59 & 60-79 & 80+ \\ \midrule
        $T_E$ & Latent period & Days & 3.335 & 3.335 & 3.335 & 3.335 & 3.335 & 3.335 \\ 
        $T_{I_{NS}}$ & Recovery time (asympt.) & Days & 1.865 & 1.865 & 1.865 & 1.865 & 1.865 & 1.865 \\ 
        $T_{I_{Sy}}$ & Recovery time (sympt.) & Days & 7.026 & 7.026 & 7.067 & 6.939 & 6.835 & 6.775 \\ 
        $T_{I_{Sev}}$ & Recovery time (severe inf.) & Days & 5.00 & 5.00 & 5.925 & 7.55 & 8.50 & 11.00 \\ 
        $T_{I_{Cr}}$ & Recovery time (critical inf.) & Days & 6.95 & 6.95 & 6.86 & 17.36 & 17.10 & 11.60 \\ 
        $\rho$ & Transm. probability & -- & 0.03 & 0.06 & 0.06 & 0.06 & 0.09 & 0.175 \\ 
        $\xi_{I_{NS}}(t)$ & Rel. infectiousness (asympt.) & -- & 1.0 & 1.0 & 1.0 & 1.0 & 1.0 & 1.0 \\ 
        $\xi_{I_{Sy}}(t)$ &  Rel. infectiousness (sympt.)  & -- & 0.3 & 0.3 & 0.3 & 0.3 & 0.3 & 0.3 \\ 
        $\mu_{I_{NS}}^{I_{Sev}}$ & Symptomatic fraction & -- & 0.75 & 0.75 & 0.8 & 0.8 & 0.8 & 0.8 \\ 
        $\mu_{I_{Sy}}^{I_{Sev}}$ & Severe fraction & -- & 0.0075 & 0.0075 & 0.019 & 0.0615 & 0.165 & 0.225 \\ 
        $\mu_{I_{Sev}}^{I_{Cr}}$ & Critical fraction & -- & 0.075 & 0.075 & 0.075 & 0.15 & 0.3 & 0.4 \\ 
        $\mu_{I_{Cr}}^{D}$ & Death probability & -- & 0.05 & 0.05 & 0.14 & 0.14 & 0.4 & 0.6 \\ 
        \bottomrule
    \end{tabular}
    \end{adjustwidth}
    \label{tab:covid_age_dep_parameters}
\end{table}

\begin{table}[htbp]
    \centering
    \caption{\textbf{Model parameters without dependency on age.} Parameters for the gamma function and the soft-plus curvature are chosen based on~\cite{10.3389/fphy.2022.842180}. The nominal ICU capacity is derived from observations during the pandemic in Germany and is further explained in the Discussion section.}
    \begin{adjustwidth}{-0.9in}{0in}
    \begin{tabular}{lp{5.3cm}lccc}
        \toprule
        Parameter & Description & Units &  Value \\ \midrule
        $a$ & Shape parameter & -- &  6  \\ 
        $b$ & Scale parameter & -- &  4 \\ 
        $\epsilon$ & Soft-plus curvature & -- & 0.1  \\ 
        $C_{ICU}$ & Nominal ICU capacity & ICU beds/100k individuals & 9    \\
        \bottomrule
    \end{tabular}
    \end{adjustwidth}
    \label{tab:covid_parameters}
\end{table}

We focus on several observables when analyzing the distribution of peaks throughout the country: i) day of peak (defined as in Sec.~\ref{sec::wave}), ii) variability on the day of peak (measured as the standard deviation of the peak day distribution), iii) peak ICU occupancy, and iv) variability of the peak value (measured as the standard deviation of the peak ICU occupancy distribution). Unless stated otherwise, we consider that all spatial levels of information have the same weights.

\subsection{Dynamic regulation of effective contacts helps de-synchronize epidemic peaks}

\begin{figure}[ht!]%
\hspace*{-2.5 cm}
        \centering
       	\includegraphics[width = 180mm]{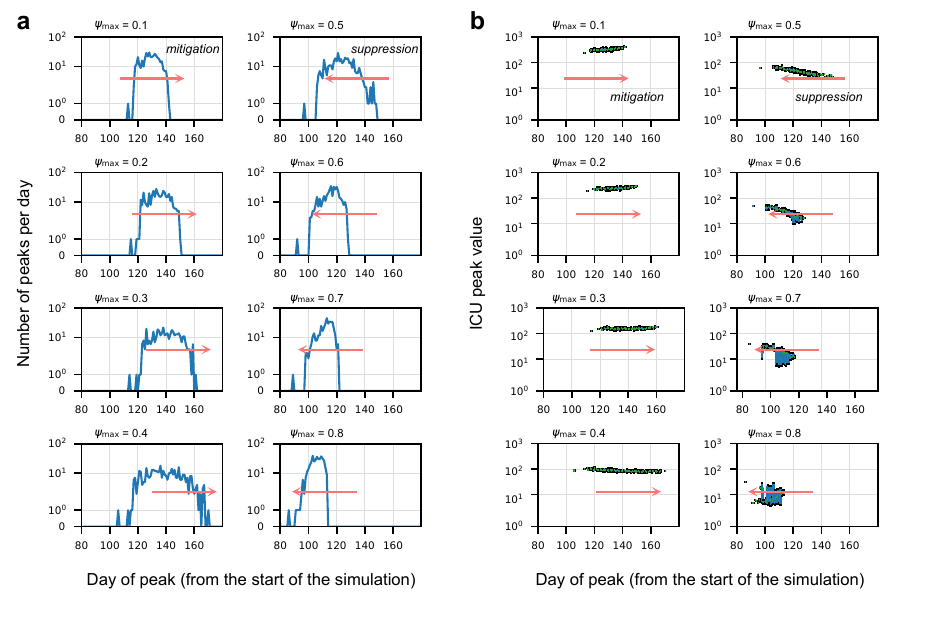}
            \caption{\textbf{Risk-mediated dynamical adaptation of effective contacts may delay the spatial spread of infectious diseases}. a) Number of ICU peaks observed, across Germany, per simulation day and b) ICU peak value per simulation day. For both panels, strengthening contact reduction from top to bottom and from bottom-most plot in 1st and 3rd column to 2nd and 4th column in steps of 0.1. As individuals become aware of infectious disease outbreaks, they may change their behavior in response to the risk they perceive. Governments respond to quantified risk metrics or these perceptions by implementing interventions that limit effective contacts, which are interactions that could lead to infection. Strengthening these interventions ($\psi_{\rm max}$) can help reduce peak occupancy in ICUs (\textbf{b}) and delay its onset until the turnover point is reached (\textbf{a}). At this point, the interventions are capable of fully extinguishing the local outbreak, allowing us to transition from mitigation to suppression.}
       	\label{fig:results_1}
\end{figure}

Our findings reveal distinct regimes that emerge as we increase the strength of the behavioral feedback, represented by $\psi_{\rm max}$, which corresponds to the highest level of contact reduction achievable through realistic self-regulation or interventions (cf.~\cref{fig:results_1}). As we increase $\psi_{\rm max}$ until $\psi_{\rm max}\approx 0.4$, we observe two primary effects on the timing and magnitude of epidemic peaks. First, there is a notable delay in the mean occurrence of peak days across regions. This delay is particularly pronounced as $\psi_{\rm max}$ increases from low to moderate values, suggesting that even modest levels of dynamic contact reduction can significantly postpone the onset of major outbreaks. Concurrently, we observe an increase in the dispersion of peak timings, indicating a de-synchronization of epidemic waves across different geographical areas. This de-synchronization could potentially alleviate the strain on healthcare systems by distributing peak demand for medical resources over a longer period.

However, as $\psi_{\rm max}$ continues to increase, we identify a transition point that marks a shift from a "mitigation" regime to a "suppression" regime. In the mitigation regime, interventions are sufficient to slow the spread of the disease but not enough to fully contain it, resulting in delayed but still significant outbreaks. The suppression regime, on the other hand, is characterized by interventions strong enough to push the effective reproduction number below 1, leading to the extinguishment of local outbreaks before they can spread widely. This transition is evidenced by a narrowing of the peak day distributions towards initial scenario values and a sharp decrease in peak ICU occupancy.

\begin{figure}[ht!]%
\hspace*{-2.5 cm}
        \centering
       	\includegraphics[width = 180mm]{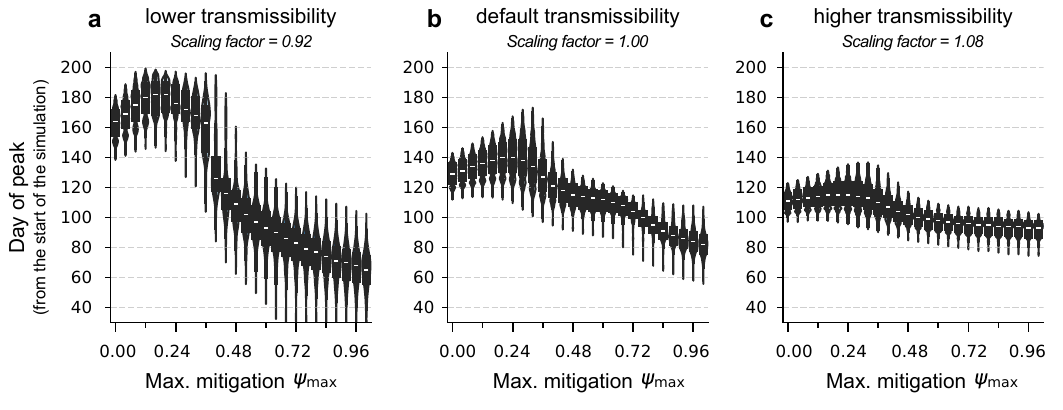}
       	\caption{\textbf{Base transmissibility of the disease determines both the dispersion and duration of the mitigation regime.} We compare the peak day distributions for different variations of our default scenario of base transmissibility (b): lower (a) and higher (c). We see that this scaling factor determines both the natural peak day distribution and the minimum feedback strength required for the transition between mitigation and suppression.}
       	\label{fig:results_2}%
\end{figure}

The range of $\psi_{\rm max}$ values corresponding to the mitigation regime strongly depends on the base transmissibility of the disease. For instance, when analyzing scenarios with lower or higher transmissibility of our default values (cf.~\cref{fig:results_2}), we see that the threshold for the transition from mitigation to suppression becomes diffuse. On the one hand, for lower transmissibility (\cref{fig:results_2}a), waves are naturally delayed, and suppressing them becomes easier; thus, the turnover occurs earlier. On the other hand, higher transmissibility requires stronger actions (\cref{fig:results_2}c). This phenomenon can be attributed to the faster progression of the epidemic wave under higher transmission rates, resulting in populations reaching final size thresholds more rapidly. Consequently, the window of opportunity for mitigation strategies narrows, and the potential for containment is strongly related to the delay associated with the calculation of $H$ (i.e., the shape and scaling parameters of the memory kernel, which have been set to default in all scenarios). Adapting this parameter, allowing for faster and even preventive interventions, could help suppress outbreaks earlier~\cite{wagner2023societal}.

In the next section, we seek to quantify this measure of dispersion. 

\subsection{Global rather than local information helps de-synchronize epidemic waves in the mitigation regime}

Up to this point, we have considered that all spatial sources of information are equally important when building the risk perception or quantification. However, when deciding on interventions, local authorities can deem the information from the immediate neighboring regions more important than the national averages. While fixing the local weight of information, we systematically explore this effect by continuously changing the weight of the regional information relative to the national information and assessing its impact in both mean and variability of the peak day distributions. 

\begin{figure}[ht!]%
\hspace*{-2.5 cm}
        \centering
       	\includegraphics[width = 180mm]{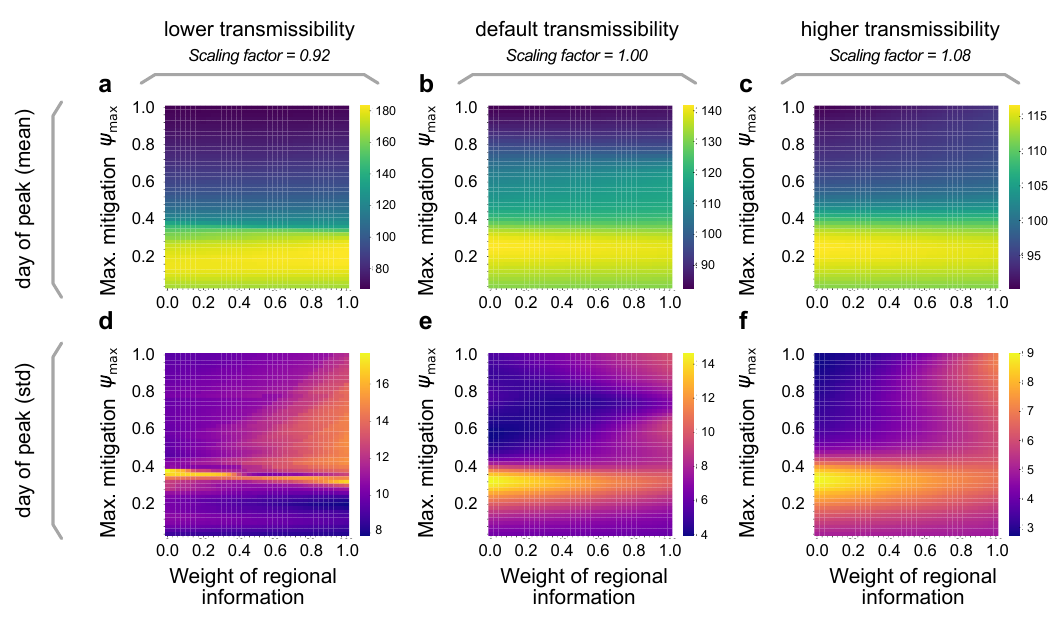}
       	\caption{\textbf{Global rather than local information helps disperse epidemic waves in the mitigation regime}. We analyze the effect of changing the weight of regional information relative to global/national information on the mean (\textbf{a--c}) and dispersion of the peak day distribution (\textbf{d--f}), for different scaling factors of the base transmissibility. We find a marked transition accounting for the mitigation-suppression turnover in the mean peak day (\textbf{a--c}), but no sensitivity to the information source. However, the dispersion of the peak day distribution responds strongly to changes in the weighting factor (\textbf{d--f}), reversing trends when transitioning from mitigation to suppression. This suggests that there may be an optimal range of $\psi_{\rm max}$ values that balance the benefits of mitigation with the costs of stringent interventions.}
       	\label{fig:results_3}%
\end{figure}

Firstly, as we move up the y-axis of~\cref{fig:results_3}a--c, representing an increase in the maximum mitigation factor ($\psi_{\rm max}$), we observe a marked transition from mitigation to suppression of epidemic waves. This transition is characterized by a shift towards earlier peak days and a reduction in the dispersion of peak timings. Rather than completely suppressing some waves, this phenomenon suggests that stronger contact reductions lead to a more rapid and synchronized response across regions. Note that the color bars have different limits for each scaling factor, given that peaks occur earlier as the disease's base transmissibility increases.

Secondly, although there is no marked effect for the mean day of peak (\cref{fig:results_3}a--c), the regional blending factor appears to influence the dispersion of epidemic waves (\cref{fig:results_3}d--f). When national information is weighted more heavily than regional information (in the mitigation regime), waves tend to be more dispersed across regions. This suggests that relying more on national-level data for risk perception and decision-making may lead to greater heterogeneity in the timing of epidemic peaks across different areas. The effect reverses when analyzing the suppression regime. The relationship between the maximum contact reduction ($\psi_{\rm max}$) and epidemic outcomes is not trivial. While increasing $\psi_{\rm max}$ does provide the system with greater potential to reduce the effective spread of the disease, the actual impact on case numbers and peak timing depends on complex interactions between behavioral changes, intervention timing, and disease transmissibility. The Figure suggests that there may be an optimal range of $\psi_{\rm max}$ values that balance the benefits of mitigation with the costs of stringent interventions.

\subsection{Political boundaries affect containment}

In the following, we investigate the impact of different spatial aggregations on our scenarios. We start by considering a maximum allowed contact reduction of $\psi_{max}=0.36$, as this value represents the transition from mitigation to suppression of the epidemic waves (see~\cref{fig:results_2}a-c) and also shows a significant difference due to the different weighting of regional and national risks (see~\cref{fig:results_3}a-f).
Therefore, we fix the impact of the integrated local risk $w^{(k)}=0.3$ while varying regional and national risks, $w^{(r)}$ and $w^{(n)}$. 

\begin{figure}[ht!]%
\hspace*{-1.0 cm}
        \centering
       	\includegraphics[width = 120mm]{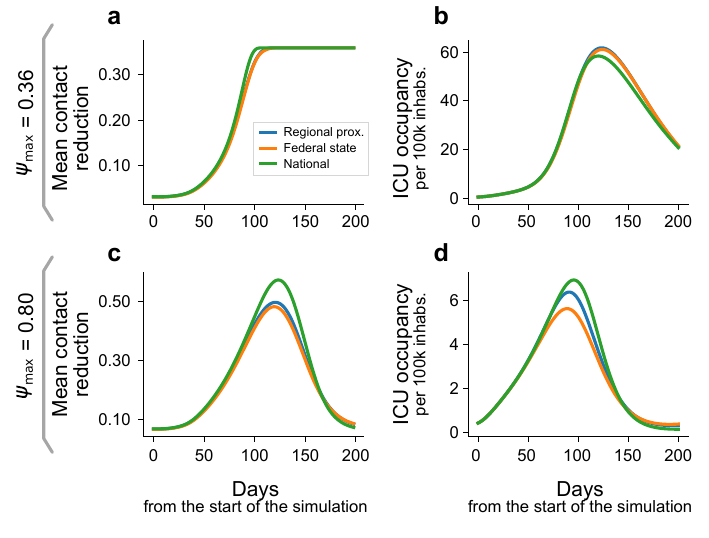}
       	\caption{\textbf{Total impact of contact reduction and ICU occupancy.} Contact reduction (a,c) and ICU occupancy per 100,000 individuals (b,d) for scenarios with $\psi_{max}=0.36$. The three panels compare the national, federal state, and regional proximity aggregations. The left panel shows the progression of the mean average contact reduction for all counties over time.  The mean average contact reduction is first calculated per county and then weighted by the local populations. The right panel displays the total ICU occupancy per 100,000 individuals. In scenarios with lower maximum contact reduction, all approaches yield broadly comparable results. However, when higher levels of contact reduction are allowed, the national-level approach leads to higher mean contact reduction across the country, whereas the two spatially targeted approaches yield slightly lower mean contact reduction and also ICU peak occupancies. }
       	\label{fig:results_34_kmax_036}%
\end{figure}

We consider three scenarios: (i) national risk only, (ii) regional risk based on federal states, and (iii) regional risk based on proximity (radius 103 km). \cref{fig:results_34_kmax_036} (top row) shows the average contact reduction (left) and total ICU occupancy (right) for each scenario. 
We observe that the maximal contact reduction is reached in all scenarios at the midpoint of the simulation period and lasts until the end. The first scenario to reach the maximal possible mean contact reduction is when considering only the national risk. All other scenarios consider only regional risks and show only slight differences in the mean contact reduction and total ICU occupancy in Germany. Overall, we observe a high ICU occupancy in all scenarios, indicating that the maximum allowed contact reduction is not sufficient.

\begin{figure}[ht!]%
\hspace*{-2.5 cm}
        \centering
            \includegraphics[width = 120mm]{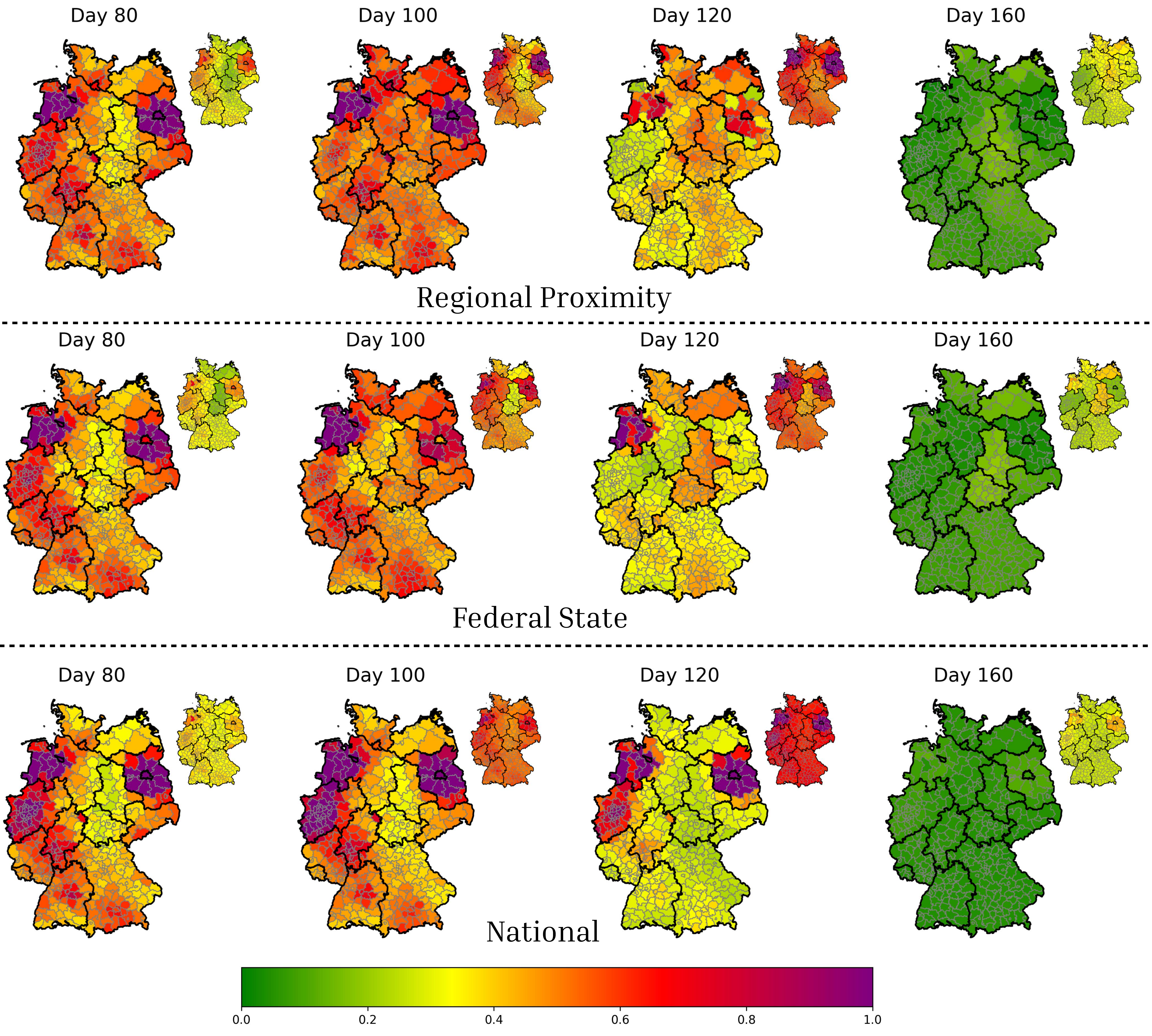}
            \caption{\textbf{Spatial distribution of ICU occupancy and risk for different scenarios at $\psi_{max}=0.8$.} The maps show the ICU occupancy on the county level and the corresponding risk (top right above) on selected days of the simulation period: Day 80, Day 100, Day 120, and Day 160. The Rows represent different aggregation levels: regional proximity (top), federal state (middle), and national (bottom). 
            ICU occupancy rates are limited to $1.0$ for better visualization. Notably, the federal state-based approach isolates certain hotspots better, leading to more localized peaks than in the regional proximity scenario. In contrast, in the national-level approach, uniform risk perception causes more regions to exceed the nominal ICU capacity due to insufficient local contact reduction.}
            \label{fig:results_34_kmax_080_maps}%
\end{figure}

To further examine regional aggregation effects, we analyze a case with a higher maximum contact reduction of $\psi_{\text{max}} = 0.8$, with results shown in~\cref{fig:results_34_kmax_036} (bottom row). Unlike the previous case, mean contact reduction does not reach the maximum in all scenarios. The national scenario has the highest mean contact reduction, while regional scenarios show similar levels.
However, ICU occupancy varies significantly. Despite the highest mean contact reduction, the national scenario results in the highest ICU usage. The federal state-based regional scenario performs best in reducing total ICU occupancy, while the regional proximity scenario leads to higher ICU levels than the federal state approach.
To explain these differences, we analyze the spatial ICU distribution in~\cref{fig:results_34_kmax_080_maps}. This figure presents ICU occupancy and perceived risk on the county level for selected simulation days. In the national scenario, perceived risk is evenly distributed, despite regional ICU differences. Lower-infection areas experience stronger contact reductions, suppressing ICU demand, but at the cost of weaker reductions in hotspots, leading to local ICU overflows. This imbalance creates widespread peaks around the hotspot regions. The federal state-based scenario results in more heterogeneous risk distribution. For instance, Berlin, located at the center of the eastern hotspot, on day 80 shows lower ICU occupancy due to its isolated risk perception, whereas in other scenarios, its risk and ICU levels align with surrounding areas. An artifact might be given in the two different regional approaches, due to the role of Berlin, also representing a disease hotspot here. It is important to note that the federal state approach prevents Berlin's ICU capacity from being exceeded, unlike with the local proximity definition where surrounding counties lower the locally present risk, reducing contact reductions and leading to a higher number of effective contacts. Due to its high population, Berlin significantly impacts many counties, even those at greater distances.

A similar effect is observed around day 80 in Lower Saxony, the federal state that hosts the hotspot in the Northwest. Here, the federal state scenario isolates the hotspot by raising regional risk, keeping neighboring states largely unaffected. In contrast, the regional proximity scenario fails to isolate the hotspot, as lower-risk neighboring counties reduce the risk fed back in Lower Saxony, leading to insufficient contact reduction and increased outbreaks in other regions.

\clearpage
\section{Discussion}

In this study, we proposed a model to investigate the relationship between the dynamic regulation of effective contacts triggered by the quantified risk that is based on the disease-related ICU occupancy and the spatial dynamics of epidemic outbreaks. Our goal was to shed light on this complex relationship and interplay using a graph-based metapopulation ODE model for disease spread, which incorporates various levels of information (local, regional, and national). We aimed to isolate the impact of this feedback loop in the metapopulation model while controlling for the maximum contact reduction that interventions can achieve and the relative weight of regional information compared to national information in influencing risk perception.

Our findings indicate that the dynamic adaptation of interventions leads to two distinct regimes depending on the maximum contact reduction, $\psi_{\rm max}$: (modest) mitigation and suppression. In the mitigation regime, where interventions are insufficient to reduce the effective spreading rate below the epidemic threshold (or fail to do so in a timely manner), gradually increasing $\psi_{\rm max}$ from zero to moderate levels delayed and spread out the onset of infection waves (in relation to their peak day), while also gradually reducing the peak values. The result of a more dispersed onset of waves is the potential to reallocate resources from less-affected areas to support those that are more impacted. In the suppression regime, the feedback-induced contact reductions can extinguish local outbreaks, thereby decreasing the dispersion of peak day distribution and shifting the mean toward the beginning of the simulation. As noted in~\cite{wagner2023societal}, the delay in the feedback response plays a critical role in determining the limits of this effect.

The best sources of information to consider depend on the overall ability to reduce effective contacts ($\psi_{\rm max}$). If the strength of interventions is only sufficient to (modestly) mitigate outbreaks, it is advisable to focus on national information when planning dynamic interventions as also stated in~\cite{ferguson_impact_2020,flaxman_estimating_2020}. This approach allows communities to prepare for the introduction of new cases and helps to flatten their infection curves. Conversely, if $\psi_{\rm max}$ is adequate to suppress local outbreaks, it is more effective to consider regional information. This way, outbreaks can be thoroughly suppressed. However, this effect is less evident when the disease's transmissibility is high, as we encounter the limitations of delayed intervention responses which can significantly undermine even strong interventions~\cite{prem_effect_2020, flaxman_estimating_2020}. The effectiveness of a local approach, also implementing wide-spread testing on commuters, was also shown in~\cite{kuhn_regional_2022}.

We analyzed a default scenario characterized by parameters reflecting the epidemic situation in Germany starting in October 2020. Throughout the COVID-19 pandemic, model requirements have continually evolved, necessitating the inclusion of vaccinations~\cite{nordstrom_risk_2022}, variants~\cite{marcelin_covid-19_2022, Nyberg2022}, and waning immunity~\cite{BOBROVITZ2023,2023833}. We examined several versions of the modeling framework (without the behavioral response), including vaccination~\cite{koslow_appropriate_2022} and waning immunity~\cite{zunker_novel_2024}. However, we focused on the simplest version of the model to better isolate the effects of behavioral feedback. 

At the beginning of the pandemic, the maximum ICU capacity in Germany was around 30,000 beds~\cite{statista2024}. However, this figure includes all types of patients, not just those with COVID-19, and the effective capacity decreased over time due to personal shortages~\cite{bundestag2021}. The peak ICU occupancy due to COVID-19 was observed in December 2020/January 2021, reaching around 21.5\% of the total ICU capacity~\cite{DIVI22, karagiannidis_case_2020}, leading to the postponement of non-emergency treatments~\cite{bundestag2021} which had further consequences that are beyond the scope of this work~\cite{10.1002/bjs.11746}. Therefore, we assume a nominal ICU capacity of 7,500 beds specifically for COVID-19 patients for our simulations, corresponding to approximately 9 ICU beds per 100,000 individuals, based on Germany's total population of 83 million. This value sets the parameter for the behavioral response in the feedback mechanism, $C_{ICU}$, which denotes the level at which mitigation measures reach their peak.

Our analysis has several limitations. First, the search space for potential parameter variations is extensive. We decided to systematically explore variations in the strength of the feedback loop and the sources of information. Sensitivity analyses regarding variations in the initialization date are provided in the supplementary materials. Second, in realistic contexts, local policies (e.g., criteria for defining risk levels) can differ from one county to another.
Additionally, we are not considering the potential effects of travelers or information from abroad here---which can be trivially included by extending the metacommunity as required.
Lastly, deterministic ODE models may not effectively capture stochasticity at low-case numbers, potentially introducing artifacts when contact reduction is excessive. However, this limitation does not affect the conclusions drawn in the regimes under consideration.

\section{Conclusion}

The desynchronization of epidemic waves allows for a more effective redistribution of resources, which can help mitigate the aftermath of an epidemic. However, achieving this is challenging. Moderate dynamic interventions that respond to outbreaks can delay the peak demand for ICU resources and increase the dispersion of these peaks. This dispersion enables better allocation of resources. If contact restrictions are only sufficient to (modestly) mitigate the spread of the disease, using national rather than regional information to shape risk perception will maximize the dispersion of epidemic waves. Conversely, if contact restrictions are strong enough to suppress the waves, it is essential to prioritize regional information in local policies.

The relationship between peak infection values and $\psi_{\rm max}$ is clear; more effective contact reduction leads to a reduction in overall infections. However, enhancing the strength or effectiveness of these interventions, as well as extending their duration, incurs higher societal costs. This creates a complex optimization challenge, as it requires balancing the goals of minimizing infections while limiting the intensity of the interventions.


\section*{Declaration of competing interest}
The authors declare that the research was conducted in the absence of any commercial or financial relationships that could be construed as a potential conflict of interest.

\section*{CRediT authorship contribution statement}
\begin{itemize}
    \item[] \textbf{Henrik Zunker:} Conceptualization, Data curation, Formal analysis, Investigation, Methodology, Software, Validation, Visualization, Writing – original draft, Writing – review \& editing 
    \item[] \textbf{Philipp Dönges:} Conceptualization, Validation, Writing – review \& editing 
    \item[] \textbf{Patrick Lenz:} Data curation, Writing – review \& editing 
    \item[] \textbf{Seba Contreras:} Conceptualization, Funding acquisition, Investigation, Methodology, Supervision, Validation, Visualization, Writing – original draft, Writing – review \& editing 
    \item[] \textbf{Martin J. Kühn:} Conceptualization, Funding acquisition, Investigation, Methodology, Project administration, Resources, Supervision, Validation, Writing – original draft, Writing – review \& editing 
\end{itemize}

\section*{Acknowledgements}
This work was supported by the Initiative and Networking Fund of the Helmholtz Association (grant agreement number KA1-Co-08, Project LOKI-Pandemics) and by the German Federal Ministry for Digital and Transport under grant agreement FKZ19F2211A (Project PANDEMOS). 
It was furthermore supported by the Max Planck Society, the German Federal Ministry for Education and Research for the RESPINOW (031L0298) and infoXpand (031L0300A) projects, and the Niedersachsen Ministry for Science and Culture  (MWK) via the programs “zukunft.niedersachsen”, “Niedersächsisches Vorab” and "Niedersachsen-Profil-Professur“ (A.Nr. 5457008).
During the preparation of this work, the authors used Grammarly AI and Perplexity AI to polish the text and enhance its readability. After utilizing these tools, the authors carefully reviewed and edited the output as necessary and take full responsibility for the final content of the published article.

\section*{Data availability}
All data and code used to generate the results presented in this work are publicly available as part of the MEmilio framework~\cite{dlr209739, kuhn_2024_14237545} on branch 978-feedback-model in the repository.





\bibliographystyle{elsarticle-num}
\bibliography{literature.bib}

\begin{thebibliography}{10}
\expandafter\ifx\csname url\endcsname\relax
  \def\url#1{\texttt{#1}}\fi
\expandafter\ifx\csname urlprefix\endcsname\relax\def\urlprefix{URL }\fi
\expandafter\ifx\csname href\endcsname\relax
  \def\href#1#2{#2} \def\path#1{#1}\fi

\bibitem{10.3389/fphy.2022.842180}
P.~Dönges, J.~Wagner, S.~Contreras, E.~N. Iftekhar, S.~Bauer, S.~B. Mohr,
  J.~Dehning, A.~Calero~Valdez, M.~Kretzschmar, M.~Mäs, K.~Nagel,
  V.~Priesemann, Interplay between risk perception, behavior, and {COVID-19}
  spread, Frontiers in Physics 10.
\newblock \href {http://dx.doi.org/10.3389/fphy.2022.842180}
  {\path{doi:10.3389/fphy.2022.842180}}.

\bibitem{Vasilis2024}
V.~Bitsouni, N.~Gialelis, V.~Tsilidis, An age-structured {SVEAIR}
  epidemiological model, Mathematical Methods in the Applied Sciences 47~(16)
  (2024) 12460--12486.
\newblock \href
  {http://arxiv.org/abs/https://onlinelibrary.wiley.com/doi/pdf/10.1002/mma.10165}
  {\path{arXiv:https://onlinelibrary.wiley.com/doi/pdf/10.1002/mma.10165}},
  \href {http://dx.doi.org/https://doi.org/10.1002/mma.10165}
  {\path{doi:https://doi.org/10.1002/mma.10165}}.

\bibitem{ploetzke2024}
L.~Plötzke, A.~Wendler, R.~Schmieding, M.~J. Kühn,
  \href{https://arxiv.org/abs/2412.09140}{Revisiting the linear chain trick in
  epidemiological models: Implications of underlying assumptions for numerical
  solutions}Submitted for publication.
\newblock \href {http://arxiv.org/abs/2412.09140} {\path{arXiv:2412.09140}}.
\newline\urlprefix\url{https://arxiv.org/abs/2412.09140}

\bibitem{Wendler2024IDE}
A.~C. Wendler, L.~Plötzke, H.~Tritzschak, M.~J. Kühn,
  \href{https://arxiv.org/abs/2408.12228}{A nonstandard numerical scheme for a
  novel {SECIR} integro differential equation-based model with nonexponentially
  distributed stay times.}Submitted for publication.
\newline\urlprefix\url{https://arxiv.org/abs/2408.12228}

\bibitem{pei_differential_2020}
S.~Pei, S.~Kandula, J.~Shaman, Differential effects of intervention timing on
  {COVID}-19 spread in the {United} {States}, Science Advances 6~(49) (2020)
  eabd6370.
\newblock \href {http://dx.doi.org/10.1126/sciadv.abd6370}
  {\path{doi:10.1126/sciadv.abd6370}}.

\bibitem{kuhn_assessment_2021}
M.~J. K\"uhn, D.~Abele, T.~Mitra, W.~Koslow, M.~Abedi, K.~Rack, M.~Siggel,
  S.~Khailaie, M.~Klitz, S.~Binder, L.~Spataro, J.~Gilg, J.~Kleinert,
  M.~Häberle, L.~Plötzke, C.~D. Spinner, M.~Stecher, X.~X. Zhu, A.~Basermann,
  M.~Meyer-Hermann, Assessment of effective mitigation and prediction of the
  spread of {SARS}-{CoV}-2 in {Germany} using demographic information and
  spatial resolution, Mathematical Biosciences (2021) 108648\href
  {http://dx.doi.org/https://doi.org/10.1016/j.mbs.2021.108648}
  {\path{doi:https://doi.org/10.1016/j.mbs.2021.108648}}.

\bibitem{kuhn_regional_2022}
M.~J. Kühn, D.~Abele, S.~Binder, K.~Rack, M.~Klitz, J.~Kleinert, J.~Gilg,
  L.~Spataro, W.~Koslow, M.~Siggel, M.~Meyer-Hermann, A.~Basermann,
  \href{https://bmcinfectdis.biomedcentral.com/articles/10.1186/s12879-022-07302-9}{Regional
  opening strategies with commuter testing and containment of new
  {SARS}-{CoV}-2 variants in {Germany}}, BMC Infectious Diseases 22~(1) (2022)
  333.
\newblock \href {http://dx.doi.org/10.1186/s12879-022-07302-9}
  {\path{doi:10.1186/s12879-022-07302-9}}.
\newline\urlprefix\url{https://bmcinfectdis.biomedcentral.com/articles/10.1186/s12879-022-07302-9}

\bibitem{chen_compliance_2021}
X.~Chen, A.~Zhang, H.~Wang, A.~Gallaher, X.~Zhu, Compliance and containment in
  social distancing: mathematical modeling of {COVID}-19 across townships,
  International Journal of Geographical Information Science 35~(3) (2021)
  446--465, publisher: Taylor \& Francis \_eprint:
  https://doi.org/10.1080/13658816.2021.1873999.
\newblock \href {http://dx.doi.org/10.1080/13658816.2021.1873999}
  {\path{doi:10.1080/13658816.2021.1873999}}.

\bibitem{levin_effects_2021}
M.~W. Levin, M.~Shang, R.~Stern, Effects of short-term travel on {COVID}-19
  spread: {A} novel {SEIR} model and case study in {Minnesota}, PLOS ONE 16~(1)
  (2021) e0245919, publisher: Public Library of Science.
\newblock \href {http://dx.doi.org/10.1371/journal.pone.0245919}
  {\path{doi:10.1371/journal.pone.0245919}}.

\bibitem{liu_modelling_2022}
J.~Liu, G.~P. Ong, V.~J. Pang,
  \href{https://linkinghub.elsevier.com/retrieve/pii/S0965856422001197}{Modelling
  effectiveness of {COVID}-19 pandemic control policies using an {Area}-based
  {SEIR} model with consideration of infection during interzonal travel},
  Transportation Research Part A: Policy and Practice 161 (2022) 25--47.
\newblock \href {http://dx.doi.org/10.1016/j.tra.2022.05.003}
  {\path{doi:10.1016/j.tra.2022.05.003}}.
\newline\urlprefix\url{https://linkinghub.elsevier.com/retrieve/pii/S0965856422001197}

\bibitem{zunker_novel_2024}
H.~Zunker, R.~Schmieding, D.~Kerkmann, A.~Schengen, S.~Diexer, R.~Mikolajczyk,
  M.~Meyer-Hermann, M.~J. Kühn,
  \href{https://dx.plos.org/10.1371/journal.pcbi.1012630}{Novel travel time
  aware metapopulation models and multi-layer waning immunity for late-phase
  epidemic and endemic scenarios}, PLOS Computational Biology 20~(12) (2024)
  e1012630.
\newblock \href {http://dx.doi.org/10.1371/journal.pcbi.1012630}
  {\path{doi:10.1371/journal.pcbi.1012630}}.
\newline\urlprefix\url{https://dx.plos.org/10.1371/journal.pcbi.1012630}

\bibitem{contreras2020multi}
S.~Contreras, H.~A. Villavicencio, D.~Medina-Ortiz, J.~P. Biron-Lattes,
  {\'A}.~Olivera-Nappa, A multi-group {SEIRA} model for the spread of
  {COVID-19} among heterogeneous populations, Chaos, Solitons \& Fractals 136
  (2020) 109925.
\newblock \href {http://dx.doi.org/10.1016/j.chaos.2020.109925}
  {\path{doi:10.1016/j.chaos.2020.109925}}.

\bibitem{bauer_relaxing_2021}
S.~Bauer, S.~Contreras, J.~Dehning, M.~Linden, E.~Iftekhar, S.~B. Mohr,
  A.~Olivera-Nappa, V.~Priesemann, Relaxing restrictions at the pace of
  vaccination increases freedom and guards against further {COVID}-19 waves,
  PLOS Computational Biology 17~(9) (2021) e1009288.
\newblock \href {http://dx.doi.org/10.1371/journal.pcbi.1009288}
  {\path{doi:10.1371/journal.pcbi.1009288}}.

\bibitem{donofrio2009information}
A.~d’Onofrio, P.~Manfredi, Information-related changes in contact patterns
  may trigger oscillations in the endemic prevalence of infectious diseases,
  Journal of Theoretical Biology 256~(3) (2009) 473--478.
\newblock \href {http://dx.doi.org/10.1016/j.jtbi.2008.10.005}
  {\path{doi:10.1016/j.jtbi.2008.10.005}}.

\bibitem{banerjee2023spatio}
M.~Banerjee, S.~Ghosh, P.~Manfredi, A.~d’Onofrio, Spatio-temporal chaos and
  clustering induced by nonlocal information and vaccine hesitancy in the sir
  epidemic model, Chaos, Solitons \& Fractals 170 (2023) 113339.
\newblock \href {http://dx.doi.org/10.1016/j.chaos.2023.113339}
  {\path{doi:10.1016/j.chaos.2023.113339}}.

\bibitem{banerjee2024behavior}
M.~Banerjee, V.~Volpert, P.~Manfredi, A.~d’Onofrio, Behavior-induced phase
  transitions with far from equilibrium patterning in a sis epidemic model:
  Global vs non-local feedback, Physica D: Nonlinear Phenomena 469 (2024)
  134316.
\newblock \href {http://dx.doi.org/10.1016/j.physd.2024.134316}
  {\path{doi:10.1016/j.physd.2024.134316}}.

\bibitem{della2021volatile}
R.~Della~Marca, A.~d’Onofrio, Volatile opinions and optimal control of
  vaccine awareness campaigns: Chaotic behaviour of the forward-backward sweep
  algorithm vs. heuristic direct optimization, Communications in Nonlinear
  Science and Numerical Simulation 98 (2021) 105768.
\newblock \href {http://dx.doi.org/10.1016/j.cnsns.2021.105768}
  {\path{doi:10.1016/j.cnsns.2021.105768}}.

\bibitem{Betsch_Wieler_Bosnjak_Ramharter_Stollorz_Omer_Korn_Sprengholz_Felgendreff_Eitze_Schmid_2020}
C.~Betsch, L.~Wieler, M.~Bosnjak, M.~Ramharter, V.~Stollorz, S.~Omer, L.~Korn,
  P.~Sprengholz, L.~Felgendreff, S.~Eitze, P.~Schmid, Germany {COVID-19}
  snapshot monitoring ({COSMO Germany}): Monitoring knowledge, risk
  perceptions, preventive behaviours, and public trust in the current
  coronavirus outbreak in germany\href
  {http://dx.doi.org/10.23668/psycharchives.2776}
  {\path{doi:10.23668/psycharchives.2776}}.

\bibitem{BETSCH20201255}
C.~Betsch, L.~H. Wieler, K.~Habersaat, Monitoring behavioural insights related
  to {COVID-19}, The Lancet 395~(10232) (2020) 1255--1256.
\newblock \href
  {http://dx.doi.org/https://doi.org/10.1016/S0140-6736(20)30729-7}
  {\path{doi:https://doi.org/10.1016/S0140-6736(20)30729-7}}.

\bibitem{wagner2023societal}
J.~Wagner, S.~Bauer, S.~Contreras, L.~Fleddermann, U.~Parlitz, V.~Priesemann,
  \href{https://arxiv.org/abs/2305.15427}{Societal feedback induces complex and
  chaotic dynamics in endemic infectious diseases}, In press in Physical Review
  Research.
\newline\urlprefix\url{https://arxiv.org/abs/2305.15427}

\bibitem{stollenwerk2022seasonally}
N.~Stollenwerk, S.~Spaziani, J.~Mar, I.~E. Arrizabalaga, D.~Knopoff,
  N.~Cusimano, V.~Anam, A.~Shrivastava, M.~Aguiar, Seasonally forced sir
  systems applied to respiratory infectious diseases, bifurcations, and chaos,
  Computational and Mathematical Methods 2022~(1) (2022) 3556043.
\newblock \href {http://dx.doi.org/10.1155/2022/3556043}
  {\path{doi:10.1155/2022/3556043}}.

\bibitem{Zozmann2024}
H.~Zozmann, L.~Sch\"{u}ler, X.~Fu, E.~Gawel, Autonomous and policy-induced
  behavior change during the covid-19 pandemic: Towards understanding and
  modeling the interplay of behavioral adaptation, PLOS ONE 19~(5) (2024)
  e0296145.
\newblock \href {http://dx.doi.org/10.1371/journal.pone.0296145}
  {\path{doi:10.1371/journal.pone.0296145}}.

\bibitem{contreras2023emergency}
S.~Contreras, E.~N. Iftekhar, V.~Priesemann, From emergency response to
  long-term management: The many faces of the endemic state of {COVID-19}, The
  Lancet Regional Health--Europe 30.
\newblock \href {http://dx.doi.org/10.1016/j.lanepe.2023.100664}
  {\path{doi:10.1016/j.lanepe.2023.100664}}.

\bibitem{kerr_covasim_2021}
C.~C. Kerr, R.~M. Stuart, D.~Mistry, R.~G. Abeysuriya, K.~Rosenfeld, G.~R.
  Hart, R.~C. Núñez, J.~A. Cohen, P.~Selvaraj, B.~Hagedorn, L.~George,
  M.~Jastrzebski, A.~S. Izzo, G.~Fowler, A.~Palmer, D.~Delport, N.~Scott, S.~L.
  Kelly, C.~S. Bennette, B.~G. Wagner, S.~T. Chang, A.~P. Oron, E.~A. Wenger,
  J.~Panovska-Griffiths, M.~Famulare, D.~J. Klein, Covasim: {An} agent-based
  model of {COVID}-19 dynamics and interventions, PLOS Computational Biology
  17~(7) (2021) 1--32, publisher: Public Library of Science.
\newblock \href {http://dx.doi.org/10.1371/journal.pcbi.1009149}
  {\path{doi:10.1371/journal.pcbi.1009149}}.

\bibitem{muller_predicting_2021}
S.~A. Müller, M.~Balmer, W.~Charlton, R.~Ewert, A.~Neumann, C.~Rakow,
  T.~Schlenther, K.~Nagel, Predicting the effects of {COVID}-19 related
  interventions in urban settings by combining activity-based modelling,
  agent-based simulation, and mobile phone data, PLOS ONE 16~(10) (2021)
  e0259037.
\newblock \href {http://dx.doi.org/10.1371/journal.pone.0259037}
  {\path{doi:10.1371/journal.pone.0259037}}.

\bibitem{KKN24}
D.~Kerkmann, S.~Korf, K.~Nguyen, D.~Abele, A.~Schengen, C.~Gerstein, J.~H.
  Göbbert, A.~Basermann, M.~J. Kühn, M.~Meyer-Hermann,
  \href{https://arxiv.org/abs/2410.08050}{Agent-based modeling for realistic
  reproduction of human mobility and contact behavior to evaluate test and
  isolation strategies in epidemic infectious disease spread}\href
  {http://arxiv.org/abs/2410.08050} {\path{arXiv:2410.08050}}.
\newline\urlprefix\url{https://arxiv.org/abs/2410.08050}

\bibitem{bmas_pendlerverflechtungen_2020}
{BMAS},
  \href{https://statistik.arbeitsagentur.de/SiteGlobals/Forms/Suche/Einzelheftsuche_Formular.html?topic_f=beschaeftigung-sozbe-krpendd}{Pendlerverflechtungen
  der sozialversicherungspflichtig {Beschäftigten} nach {Kreisen} -
  {Deutschland} ({Jahreszahlen})}, publication Title: www.bmas.de (2020).
\newline\urlprefix\url{https://statistik.arbeitsagentur.de/SiteGlobals/Forms/Suche/Einzelheftsuche_Formular.html?topic_f=beschaeftigung-sozbe-krpendd}

\bibitem{kuhn_vorlaufige_2022}
M.~Kühn, A.~Schengen, T.~Mocanu,
  \href{https://mcloud.de/en/web/guest/suche/-/results/detail/d45ae31c-6725-4e64-8ebe-51421a02ee0c}{Vorläufige
  bundesweite {Verkehrsströme}} (Dec. 2022).
\newline\urlprefix\url{https://mcloud.de/en/web/guest/suche/-/results/detail/d45ae31c-6725-4e64-8ebe-51421a02ee0c}

\bibitem{prem_projecting_2017}
K.~Prem, A.~R. Cook, M.~Jit, Projecting social contact matrices in 152
  countries using contact surveys and demographic data, PLoS computational
  biology 13~(9) (2017) e1005697.
\newblock \href {http://dx.doi.org/10.1371/journal.pcbi.1005697}
  {\path{doi:10.1371/journal.pcbi.1005697}}.

\bibitem{fumanelli_inferring_2012}
L.~Fumanelli, M.~Ajelli, P.~Manfredi, A.~Vespignani, S.~Merler, Inferring the
  {Structure} of {Social} {Contacts} from {Demographic} {Data} in the
  {Analysis} of {Infectious} {Diseases} {Spread}, PLoS Computational Biology
  8~(9) (2012) e1002673.
\newblock \href {http://dx.doi.org/10.1371/journal.pcbi.1002673}
  {\path{doi:10.1371/journal.pcbi.1002673}}.

\bibitem{Robert_Koch-Institut_SARS-CoV-2_Infektionen_in_2025}
{Robert Koch-Institut},
  \href{https://robert-koch-institut.github.io/SARS-CoV-2-Infektionen_in_Deutschland}{{SARS-CoV-2
  Infektionen in Deutschland}} (Feb. 2025).
\newblock \href {http://dx.doi.org/10.5281/zenodo.14822865}
  {\path{doi:10.5281/zenodo.14822865}}.
\newline\urlprefix\url{https://robert-koch-institut.github.io/SARS-CoV-2-Infektionen_in_Deutschland}

\bibitem{DIVI22}
{Deutsche Interdisziplin\"are Vereinigung f\"ur Intensiv- und Notfallmedizin
  (DIVI)},
  \href{https://www.divi.de/divi-intensivregister-tagesreport-archiv}{{DIVI}
  {Intensivregister} {Tagesreport}} (2022).
\newline\urlprefix\url{https://www.divi.de/divi-intensivregister-tagesreport-archiv}

\bibitem{RKIVACC}
\href{https://www.rki.de/DE/Content/InfAZ/N/Neuartiges_Coronavirus/Daten/Impfquoten-Tab.html}{Digitales
  {I}mpfquotenmonitoring zur {COVID}-19-{I}mpfung} (2023).
\newline\urlprefix\url{https://www.rki.de/DE/Content/InfAZ/N/Neuartiges_Coronavirus/Daten/Impfquoten-Tab.html}

\bibitem{ferguson_impact_2020}
N.~M. Ferguson, D.~Laydon, G.~Nedjati-Gilani, N.~Imai, K.~Ainslie, M.~Baguelin,
  S.~Bhatia, A.~Boonyasiri, Z.~Cucunuba, G.~Cuomo-Dannenburg, A.~Dighe,
  I.~Dorigatti, H.~Fu, K.~Gaythorpe, W.~Green, A.~Hamlet, W.~Hinsley, L.~C.
  Okell, S.~v. Elsland, H.~Thompson, R.~Verity, E.~Volz, H.~Wang, Y.~Wang,
  P.~G. Walker, C.~Walters, P.~Winskill, C.~Whittaker, C.~A. Donnelly,
  S.~Riley, A.~C. Ghani,
  \href{https://www.gov.uk/government/publications/impact-of-non-pharmaceutical-interventions-npis-to-reduce-covid-19-mortality-and-healthcare-demand-16-march-2020}{Impact
  of non-pharmaceutical interventions ({NPIs}) to reduce {COVID}-19 mortality
  and healthcare demand}, Tech. rep., Imperial College (2020).
\newline\urlprefix\url{https://www.gov.uk/government/publications/impact-of-non-pharmaceutical-interventions-npis-to-reduce-covid-19-mortality-and-healthcare-demand-16-march-2020}

\bibitem{flaxman_estimating_2020}
S.~Flaxman, S.~Mishra, A.~Gandy, H.~J.~T. Unwin, T.~A. Mellan, H.~Coupland,
  C.~Whittaker, H.~Zhu, T.~Berah, J.~W. Eaton, M.~Monod, {Imperial College
  COVID-19 Response Team}, P.~N. Perez-Guzman, N.~Schmit, L.~Cilloni, K.~E.~C.
  Ainslie, M.~Baguelin, A.~Boonyasiri, O.~Boyd, L.~Cattarino, L.~V. Cooper,
  Z.~Cucunubá, G.~Cuomo-Dannenburg, A.~Dighe, B.~Djaafara, I.~Dorigatti, S.~L.
  van Elsland, R.~G. FitzJohn, K.~A.~M. Gaythorpe, L.~Geidelberg, N.~C.
  Grassly, W.~D. Green, T.~Hallett, A.~Hamlet, W.~Hinsley, B.~Jeffrey,
  E.~Knock, D.~J. Laydon, G.~Nedjati-Gilani, P.~Nouvellet, K.~V. Parag,
  I.~Siveroni, H.~A. Thompson, R.~Verity, E.~Volz, C.~E. Walters, H.~Wang,
  Y.~Wang, O.~J. Watson, P.~Winskill, X.~Xi, P.~G.~T. Walker, A.~C. Ghani,
  C.~A. Donnelly, S.~Riley, M.~A.~C. Vollmer, N.~M. Ferguson, L.~C. Okell,
  S.~Bhatt, Estimating the effects of non-pharmaceutical interventions on
  {COVID}-19 in {Europe}, Nature 584~(7820) (2020) 257--261.
\newblock \href {http://dx.doi.org/10.1038/s41586-020-2405-7}
  {\path{doi:10.1038/s41586-020-2405-7}}.

\bibitem{prem_effect_2020}
K.~Prem, Y.~Liu, T.~W. Russell, A.~J. Kucharski, R.~M. Eggo, N.~Davies,
  S.~Flasche, S.~Clifford, C.~A.~B. Pearson, J.~D. Munday, S.~Abbott, H.~Gibbs,
  A.~Rosello, B.~J. Quilty, T.~Jombart, F.~Sun, C.~Diamond, A.~Gimma, K.~v.
  Zandvoort, S.~Funk, C.~I. Jarvis, W.~J. Edmunds, N.~I. Bosse, J.~Hellewell,
  M.~Jit, P.~Klepac, The effect of control strategies to reduce social mixing
  on outcomes of the {COVID}-19 epidemic in {Wuhan}, {China}: a modelling
  study, The Lancet Public Health 5~(5) (2020) e261--e270.
\newblock \href {http://dx.doi.org/10.1016/S2468-2667(20)30073-6}
  {\path{doi:10.1016/S2468-2667(20)30073-6}}.

\bibitem{nordstrom_risk_2022}
P.~Nordström, M.~Ballin, A.~Nordström, Risk of infection, hospitalisation,
  and death up to 9 months after a second dose of {COVID}-19 vaccine: a
  retrospective, total population cohort study in {Sweden}, The Lancet
  399~(10327) (2022) 814--823.
\newblock \href {http://dx.doi.org/10.1016/S0140-6736(22)00089-7}
  {\path{doi:10.1016/S0140-6736(22)00089-7}}.

\bibitem{marcelin_covid-19_2022}
J.~R. Marcelin, A.~Pettifor, H.~Janes, E.~R. Brown, J.~G. Kublin, K.~E.
  Stephenson, {COVID}-19 {Vaccines} and {SARS}-{CoV}-2 {Transmission} in the
  {Era} of {New} {Variants}: {A} {Review} and {Perspective}, Open Forum
  Infectious Diseases 9~(5) (2022) ofac124.
\newblock \href {http://dx.doi.org/10.1093/ofid/ofac124}
  {\path{doi:10.1093/ofid/ofac124}}.

\bibitem{Nyberg2022}
T.~Nyberg, N.~M. Ferguson, S.~G. Nash, H.~H. Webster, S.~Flaxman, N.~Andrews,
  W.~Hinsley, J.~L. Bernal, M.~Kall, S.~Bhatt, P.~Blomquist, A.~Zaidi, E.~Volz,
  N.~A. Aziz, K.~Harman, S.~Funk, S.~Abbott, R.~Hope, A.~Charlett, M.~Chand,
  A.~C. Ghani, S.~R. Seaman, G.~Dabrera, D.~De~Angelis, A.~M. Presanis,
  S.~Thelwall, T.~Nyberg, N.~M. Ferguson, S.~G. Nash, H.~H. Webster,
  S.~Flaxman, N.~Andrews, W.~Hinsley, J.~Lopez~Bernal, M.~Kall, S.~Bhatt,
  P.~Blomquist, A.~Zaidi, E.~Volz, N.~Abdul~Aziz, K.~Harman, S.~Funk,
  S.~Abbott, R.~Hope, A.~Charlett, M.~Chand, A.~C. Ghani, S.~R. Seaman,
  G.~Dabrera, D.~De~Angelis, A.~M. Presanis, S.~Thelwall, Comparative analysis
  of the risks of hospitalisation and death associated with {SARS-CoV-2}
  omicron (b.1.1.529) and delta (b.1.617.2) variants in {England}: a cohort
  study, The Lancet 399~(10332) (2022) 1303–1312.
\newblock \href {http://dx.doi.org/10.1016/s0140-6736(22)00462-7}
  {\path{doi:10.1016/s0140-6736(22)00462-7}}.

\bibitem{BOBROVITZ2023}
N.~Bobrovitz, H.~Ware, X.~Ma, Z.~Li, R.~Hosseini, C.~Cao, A.~Selemon,
  M.~Whelan, Z.~Premji, H.~Issa, B.~Cheng, L.~J. {Abu Raddad}, D.~L.
  Buckeridge, M.~D. {Van Kerkhove}, V.~Piechotta, M.~M. Higdon,
  A.~Wilder-Smith, I.~Bergeri, D.~R. Feikin, R.~K. Arora, M.~K. Patel,
  L.~Subissi, Protective effectiveness of previous {SARS-CoV-2} infection and
  hybrid immunity against the omicron variant and severe disease: a systematic
  review and meta-regression, The Lancet Infectious Diseases\href
  {http://dx.doi.org/https://doi.org/10.1016/S1473-3099(22)00801-5}
  {\path{doi:https://doi.org/10.1016/S1473-3099(22)00801-5}}.

\bibitem{2023833}
C.~Stein, H.~Nassereldine, R.~J.~D. Sorensen, J.~O. Amlag, C.~Bisignano,
  S.~Byrne, E.~Castro, K.~Coberly, J.~K. Collins, J.~Dalos, F.~Daoud, A.~Deen,
  E.~Gakidou, J.~R. Giles, E.~N. Hulland, B.~M. Huntley, K.~E. Kinzel,
  R.~Lozano, A.~H. Mokdad, T.~Pham, D.~M. Pigott, R.~C. {Reiner Jr.}, T.~Vos,
  S.~I. Hay, C.~J.~L. Murray, S.~S. Lim, Past sars-cov-2 infection protection
  against re-infection: a systematic review and meta-analysis, The Lancet
  401~(10379) (2023) 833--842.
\newblock \href
  {http://dx.doi.org/https://doi.org/10.1016/S0140-6736(22)02465-5}
  {\path{doi:https://doi.org/10.1016/S0140-6736(22)02465-5}}.

\bibitem{koslow_appropriate_2022}
W.~Koslow, M.~J. Kühn, S.~Binder, M.~Klitz, D.~Abele, A.~Basermann,
  M.~Meyer-Hermann, Appropriate relaxation of non-pharmaceutical interventions
  minimizes the risk of a resurgence in {SARS}-{CoV}-2 infections in spite of
  the {Delta} variant, PLOS Computational Biology 18~(5) (2022) e1010054.
\newblock \href {http://dx.doi.org/10.1371/journal.pcbi.1010054}
  {\path{doi:10.1371/journal.pcbi.1010054}}.

\bibitem{statista2024}
Statista,
  \href{https://de.statista.com/statistik/daten/studie/1246685/umfrage/auslastung-von-intensivbetten-in-deutschland/}{Auslastung
  von intensivbetten in deutschland}, accessed: 2024-08-28 (2024).
\newline\urlprefix\url{https://de.statista.com/statistik/daten/studie/1246685/umfrage/auslastung-von-intensivbetten-in-deutschland/}

\bibitem{bundestag2021}
{Wissenschaftliche Dienste des Deutschen Bundestages},
  \href{https://www.bundestag.de/resource/blob/852948/4e2fa2188808130fde49b064e0fbffbb/WD-9-052-21-pdf-data.pdf}{Behandlungskapazitäten
  in der intensivmedizinischen versorgung während der covid-19-pandemie},
  accessed: 2024-08-28 (2021).
\newline\urlprefix\url{https://www.bundestag.de/resource/blob/852948/4e2fa2188808130fde49b064e0fbffbb/WD-9-052-21-pdf-data.pdf}

\bibitem{karagiannidis_case_2020}
C.~Karagiannidis, C.~Mostert, C.~Hentschker, T.~Voshaar, J.~Malzahn,
  G.~Schillinger, J.~Klauber, U.~Janssens, G.~Marx, S.~Weber-Carstens,
  S.~Kluge, M.~Pfeifer, L.~Grabenhenrich, T.~Welte, R.~Busse, Case
  characteristics, resource use, and outcomes of 10 021 patients with
  {COVID}-19 admitted to 920 {German} hospitals: an observational study, The
  Lancet Respiratory Medicine 8~(9) (2020) 853--862.
\newblock \href {http://dx.doi.org/10.1016/S2213-2600(20)30316-7}
  {\path{doi:10.1016/S2213-2600(20)30316-7}}.

\bibitem{10.1002/bjs.11746}
{COVIDSurg Collaborative}, Elective surgery cancellations due to the {COVID-19}
  pandemic: global predictive modelling to inform surgical recovery plans,
  British Journal of Surgery 107~(11) (2020) 1440--1449.
\newblock \href
  {http://arxiv.org/abs/https://academic.oup.com/bjs/article-pdf/107/11/1440/46389233/bjs11746.pdf}
  {\path{arXiv:https://academic.oup.com/bjs/article-pdf/107/11/1440/46389233/bjs11746.pdf}},
  \href {http://dx.doi.org/10.1002/bjs.11746} {\path{doi:10.1002/bjs.11746}}.

\bibitem{dlr209739}
M.~J. K{\"u}hn, D.~Abele, D.~Kerkmann, S.~A. Korf, H.~Zunker, A.~C. Wendler,
  J.~Bicker, K.~Nguyen, R.~Schmieding, L.~Pl{\"o}tzke, P.~Lenz, M.~F. Betz,
  C.~Gerstein, A.~Schmidt, R.~Hannemann-Tamas, N.~Wa{\ss}muth, P.~Johannssen,
  H.~Tritzschak, D.~Richter, M.~Klitz, W.~Koslow, S.~Binder, M.~Siggel,
  J.~Kleinert, K.~Rack, A.~Lutz, M.~Meyer-Hermann,
  \href{https://elib.dlr.de/209739/}{Memilio v1.3.0 - a high performance
  modular epidemics simulation software} (November 2024).
\newline\urlprefix\url{https://elib.dlr.de/209739/}

\bibitem{kuhn_2024_14237545}
M.~J. Kühn, D.~Abele, D.~Kerkmann, S.~Korf, H.~Zunker, A.~Wendler, J.~Bicker,
  K.~Nguyen, R.~Schmieding, L.~Plötzke, P.~Lenz, M.~Betz, C.~Gerstein,
  A.~Schmidt, R.~Hannemann-Tamas, N.~Waßmuth, P.~Johannssen, H.~Tritzschak,
  D.~Richter, M.~Klitz, W.~Koslow, S.~Binder, M.~Siggel, J.~Kleinert, K.~Rack,
  A.~Lutz, M.~Meyer-Hermann, {MEmilio} v1.3.0 - a high performance modular
  epidemics simulation software (Nov. 2024).
\newblock \href {http://dx.doi.org/10.5281/zenodo.14237545}
  {\path{doi:10.5281/zenodo.14237545}}.

\end{thebibliography}







\end{document}








\appendix
\section{Initialization Date September 1st, 2020}
\begin{figure}[H]
  \centering
  \includegraphics[width=\textwidth]{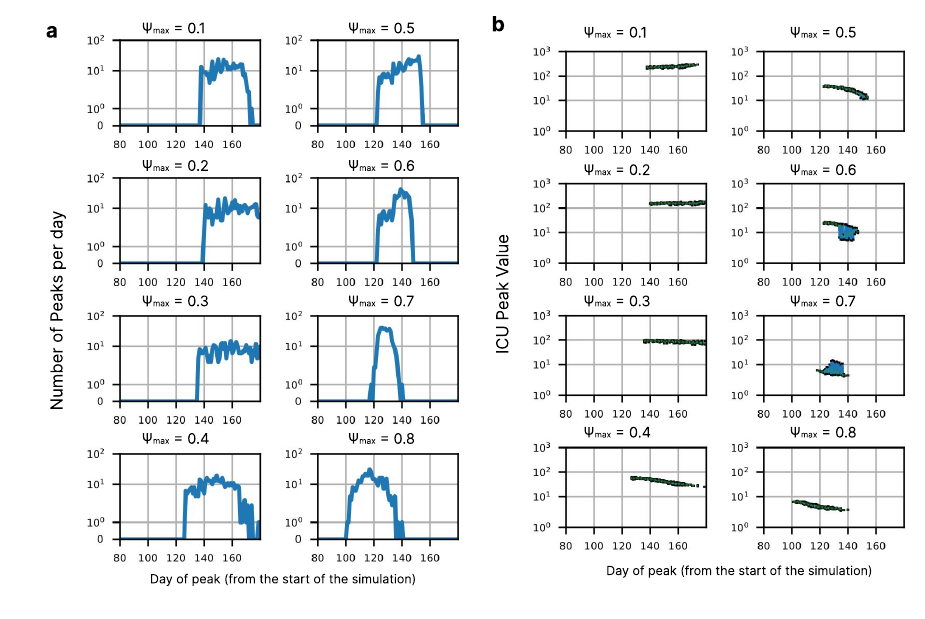}
  \caption{\textbf{Risk-mediated dynamical adaptation of interventions for initialization on September 1st, 2020.}}

  \label{fig:sep1}
\end{figure}
\begin{figure}[H]
  \centering
  \includegraphics[width=\textwidth]{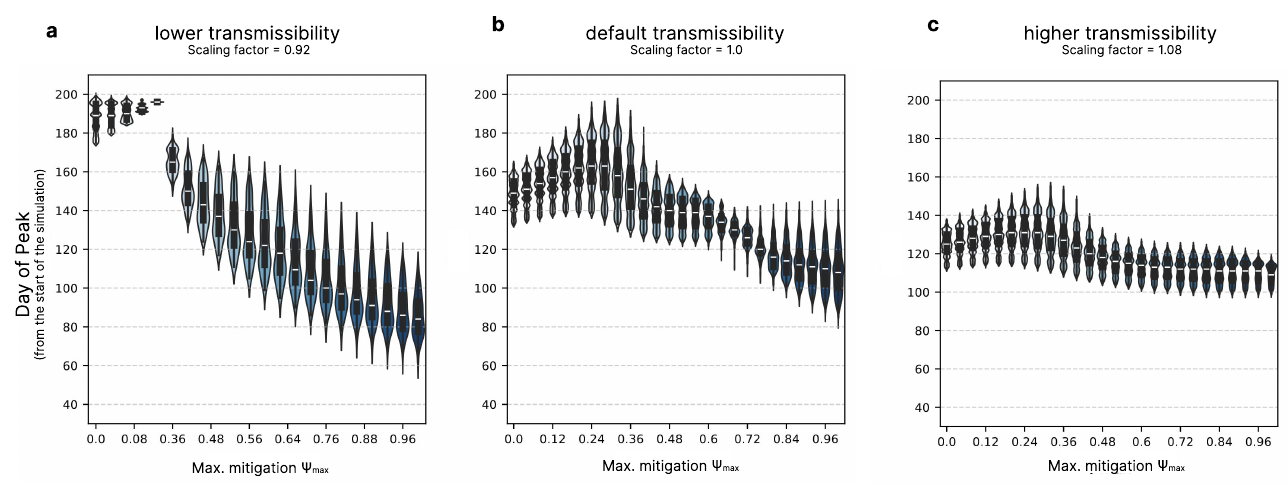}
  \caption{\textbf{Base transmissibility determines both the dispersion and duration of the mitigation regime for initialization on September 1st, 2020.}}
  \label{fig:sep2}
\end{figure}

\section{Initialization Date November 1st, 2020}
\begin{figure}[H]
  \centering
  \includegraphics[width=\textwidth]{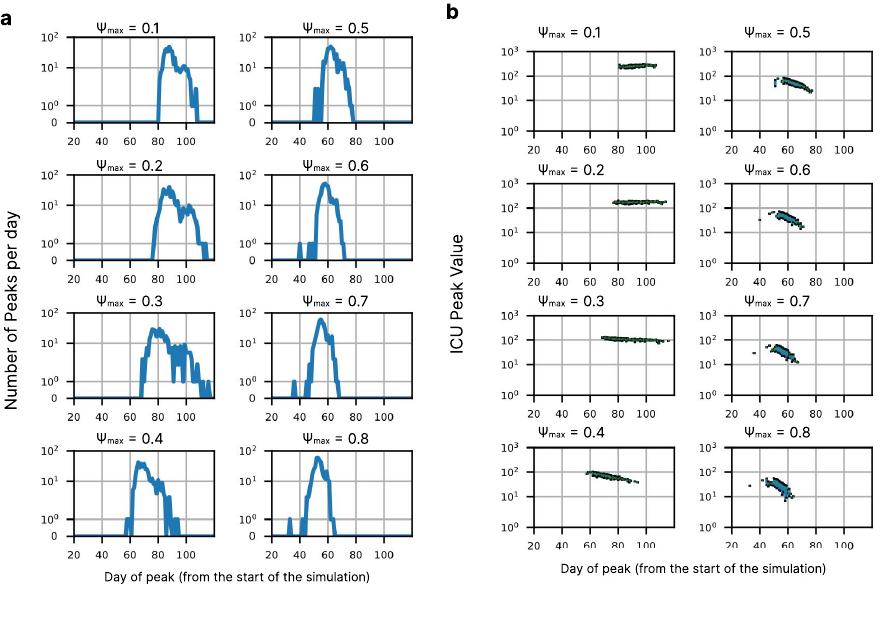}
  \caption{\textbf{Risk-mediated dynamical adaptation of interventions for initialization on November 1st, 2020.}}
  \label{fig:nov1}
\end{figure}
\begin{figure}[H]
  \centering
  \includegraphics[width=\textwidth]{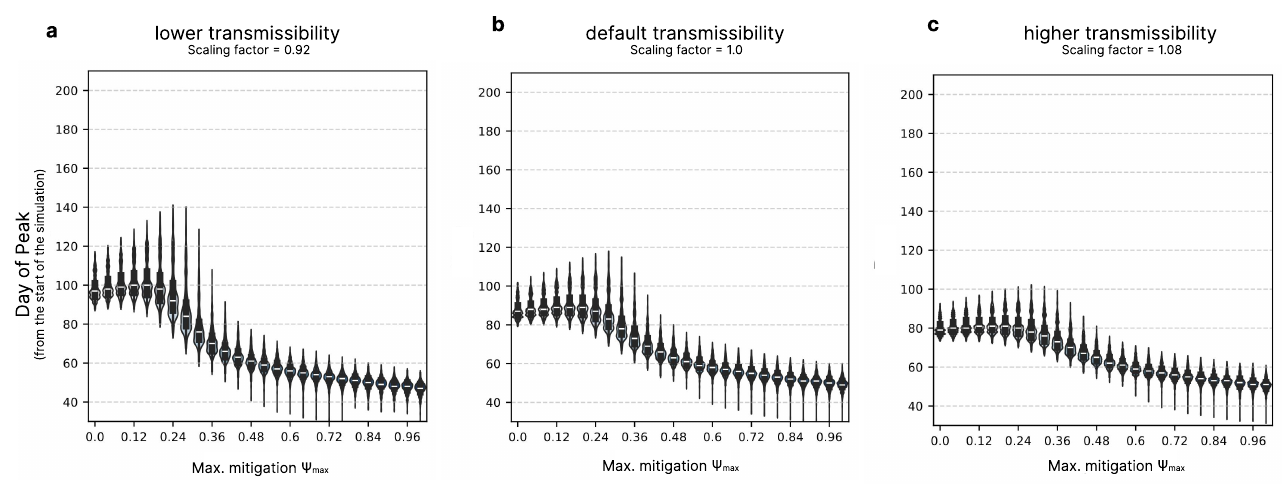}
  \caption{\textbf{Base transmissibility determines both the dispersion and duration of the mitigation regime for initialization on November 1st, 2020.}}
  \label{fig:nov2}
\end{figure}

\section{Initialization Date December 1st, 2020}
\begin{figure}[H]
  \centering
  \includegraphics[width=\textwidth]{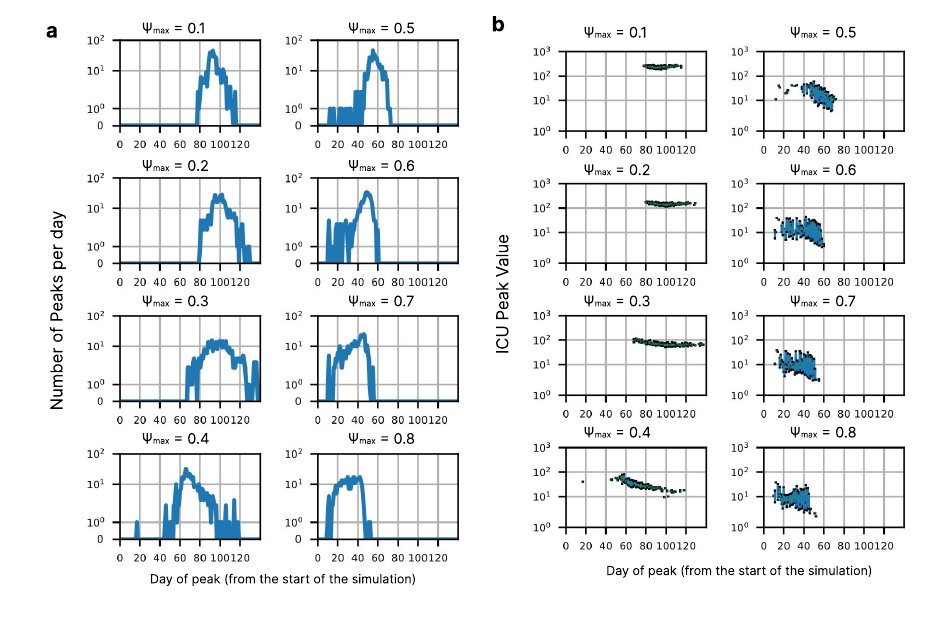}
  \caption{\textbf{Risk-mediated dynamical adaptation of interventions for initialization on December 1st, 2020.}}
  \label{fig:dec1}
\end{figure}
\begin{figure}[H]
  \centering
  \includegraphics[width=\textwidth]{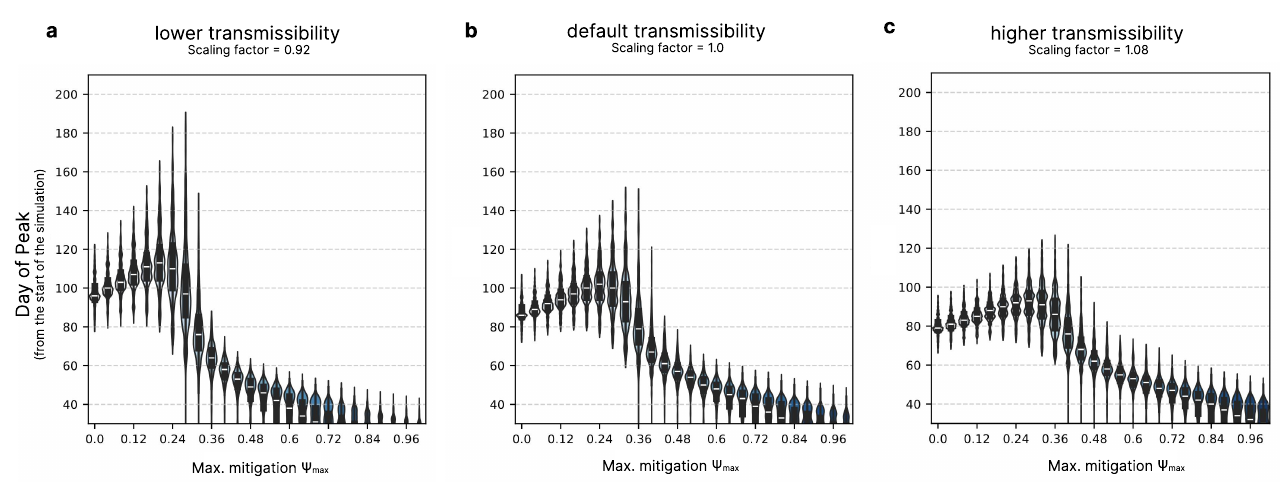}
  \caption{\textbf{Base transmissibility determines both the dispersion and duration of the mitigation regime for initialization on December 1st, 2020.}}
  \label{fig:dec2}
\end{figure}










